\documentclass[twocolumn,citeautoscript,superscriptaddress,8pt]{revtex4-2}

\usepackage[T1]{fontenc}
\usepackage{graphicx}
\usepackage{gensymb}
\usepackage{color}
\usepackage[dvipsnames]{xcolor}
\usepackage{float}
\usepackage{upgreek}
\usepackage{bm}
\usepackage{amsmath,amssymb}
\usepackage{tabularx}
\usepackage{multirow}
\usepackage{array}
\usepackage{hyperref}
\usepackage{xfrac}
\usepackage{booktabs}

\newcommand{\VB}{\ensuremath{V_{\rm B}^{-}}}
\newcommand{\dd}{$^{\ddagger}$}
\newcommand{\pdd}{\phantom{$^{\ddagger}$}}
\newcommand{\po}{\phantom{1}}
\newcommand{\ddd}{$^{\S}$}

\newcommand{\SI}{SI}

\newcolumntype{Y}{>{\centering\arraybackslash}X}

\hypersetup{urlcolor=blue, colorlinks=true, linkcolor=blue, citecolor=blue, linktocpage=true}

\begin{document}

\title{Optically detected magnetic resonance of wafer-scale hexagonal boron nitride thin films}


\author{Sam\,C.\,Scholten}
\email{s.scholten@uq.edu.au}
\thanks{Present address: School of Mathematics and Physics, University of Queensland, St.\,Lucia, Queensland 4072, Australia}
\affiliation{School of Science, RMIT University, Melbourne, VIC 3001, Australia}

\author{Jakub\,Iwański}
\affiliation{Faculty of Physics, University of Warsaw, Pasteura 5, 02-093 Warsaw, Poland}

\author{Kaijian\,Xing}
\affiliation{School of Science, RMIT University, Melbourne, VIC 3001, Australia}

\author{Johannes\,Binder}
\affiliation{Faculty of Physics, University of Warsaw, Pasteura 5, 02-093 Warsaw, Poland}

\author{Aleksandra\,K.\,Dąbrowska}
\affiliation{Faculty of Physics, University of Warsaw, Pasteura 5, 02-093 Warsaw, Poland}

\author{Hark\,H.\,Tan}
\affiliation{ARC Centre of Excellence for Transformative Meta-Optical Systems, Department of Electronic Materials Engineering, Research School of Physics, The Australian National University, Canberra, ACT 2600, Australia}

\author{Tin\,S.\,Cheng}
\affiliation{School of Physics and Astronomy, University of Nottingham, Nottingham, NG7 2RD, U.K.}

\author{Jonathan\,Bradford}
\affiliation{School of Physics and Astronomy, University of Nottingham, Nottingham, NG7 2RD, U.K.}

\author{Christopher\,J.\,Mellor}
\affiliation{School of Physics and Astronomy, University of Nottingham, Nottingham, NG7 2RD, U.K.}
    
\author{Peter\,H.\,Beton}
\affiliation{School of Physics and Astronomy, University of Nottingham, Nottingham, NG7 2RD, U.K.}

\author{Sergei\,V.\,Novikov}
\affiliation{School of Physics and Astronomy, University of Nottingham, Nottingham, NG7 2RD, U.K.}

\author{Jan\,Mischke}
\affiliation{AIXTRON SE, Dornkaulstraße 2, 52134 Herzogenrath, Germany}

\author{Sergej\,Pasko}
\affiliation{AIXTRON SE, Dornkaulstraße 2, 52134 Herzogenrath, Germany}

\author{Emre\,Yengel}
\affiliation{AIXTRON SE, Dornkaulstraße 2, 52134 Herzogenrath, Germany}

\author{Alexander\,Henning}
\affiliation{AIXTRON SE, Dornkaulstraße 2, 52134 Herzogenrath, Germany}

\author{Simonas\,Krotkus}
\affiliation{AIXTRON SE, Dornkaulstraße 2, 52134 Herzogenrath, Germany}

\author{Andrzej\,Wysmołek}
\affiliation{Faculty of Physics, University of Warsaw, Pasteura 5, 02-093 Warsaw, Poland}

\author{Jean-Philippe\,Tetienne}
\affiliation{School of Science, RMIT University, Melbourne, VIC 3001, Australia}


\begin{abstract} 
Hexagonal boron nitride (hBN) has recently been shown to host native defects exhibiting optically detected magnetic resonance (ODMR) with applications in nanoscale magnetic sensing and imaging. 
To advance these applications, deposition methods to create wafer-scale hBN films with controlled thicknesses are desirable, but a systematic study of the ODMR properties of the resultant films is lacking. 
Here we perform ODMR measurements of thin films (3-2000\,nm thick) grown via three different methods: metal-organic chemical vapour deposition (MOCVD), chemical vapour deposition (CVD), and molecular beam epitaxy (MBE). 
We find that they all exhibit an ODMR response, including the thinnest 3\,nm film, albeit with different characteristics.
The best volume-normalised magnetic sensitivity obtained is 30\,$\upmu$T\,Hz$^{-1/2}\upmu$m$^{3/2}$.
We study the effect of growth temperature on a series of MOCVD samples grown under otherwise fixed conditions and find 800-900$^\circ$C to be an optimum range for magnetic sensitivity, with a significant improvement (up to two orders of magnitude) from post-growth annealing.
This work provides a useful baseline for the magnetic sensitivity of hBN thin films deposited via standard methods and informs the feasibility of future sensing applications.   
\end{abstract}

\maketitle

\section{Introduction}

The recent discovery of optically active spin defects in hexagonal boron nitride (hBN)~\cite{tetienneQuantumSensorsGo2021} and other two-dimensional materials~\cite{liuExperimentalObservationSpin2024} has generated excitement in the colour centre community due to the unique attributes of the van der Waals (vdW) structure~\cite{aharonovichQuantumEmittersHexagonal2022}.
In sensing, the prospects for atomic-scale sample-sensor proximity, and simple integration via exfoliation and stacking into heterostructures~\cite{healeyQuantumMicroscopyVan2022}, sidestep issues encountered in engineering proximal nitrogen-vacancy (NV) ensembles in diamond~\cite{ghiasiNitrogenvacancyMagnetometryCrSBr2023, abrahamsIntegratedWidefieldProbe2021}. 
There is also interest in using 2D-confined dipolar-coupled spins for quantum simulation investigations of many-body effects in two dimensions~\cite{yaoQuantumDipolarSpin2018, davisProbingManybodyDynamics2023, abaninColloquiumManybodyLocalization2019, davisProbingManybodyDynamics2023}.
Key to development of these applications are reliable and scalable methods for creating high quality spin ensembles in thin large-area substrates.

Initial hBN sensing demonstrations~\cite{gottschollSpinDefectsHBN2021, healeyQuantumMicroscopyVan2022, huangWideFieldImaging2022, kumarMagneticImagingSpin2022, gongCoherentDynamicsStrongly2022, curieCorrelativeNanoscaleImaging2022, lyuStrainQuantumSensing2022, yangSpinDefectsHexagonal2022} utilising the boron vacancy (\VB) defect found its practicality limited by the system's low quantum efficiency, motivating a search for new spin-active colour centres.
A new class of brighter defects has been recently reported by a few groups~\cite{guoCoherentControlUltrabright2023, chejanovskySinglespinResonanceVan2021, patelRoomTemperatureDynamics2023, sternQuantumCoherentSpin2023, sternRoomtemperatureOpticallyDetected2022, mendelsonIdentifyingCarbonSource2021}, ubiquitous across growth techniques~\cite{scholtenMultispeciesOpticallyAddressable2023}, though as-yet unidentified.
Their wide emission range ($\approx400-1000$\,nm)~\cite{singhVioletNearInfraredOptical2025}, and  spread of optical and spin characteristics (including observation of various $S=1$ transitions with different zero-field splittings and isotropies~\cite{gaoSingleNuclearSpin2024, whitefieldGenerationNarrowbandQuantum2025, gilardoniSingleSpinHexagonal2024, sternQuantumCoherentSpin2023}) indicates a range of atomic structures.  
Evidence suggests a carbon-related atomic composition~\cite{mendelsonIdentifyingCarbonSource2021}, as well as a displaced spin pair nature~\cite{robertsonUniversalMechanismOptically2024}; we will presently term these systems `spin pairs'.
The common property across these spin pairs is an $S=\sfrac{1}{2}$ optically-detected magnetic resonance (ODMR) transition, which is novel amongst colour centres as being isotropic to the magnetic field orientation.
In contrast with the well-understood \VB\ defect, the spin pairs are often found at significant densities (measurable ensembles) in as-grown crystals.
However there is as-yet no known method to engineer spin-active ensembles in a controlled manner, for example tailored to specific subclasses such as those displaying an $S=1$ ground state~\cite{gaoSingleNuclearSpin2024, whitefieldGenerationNarrowbandQuantum2025, gilardoniSingleSpinHexagonal2024, sternQuantumCoherentSpin2023}.

\begin{figure*}[ht!]
\centering
\includegraphics[width=0.99\textwidth]{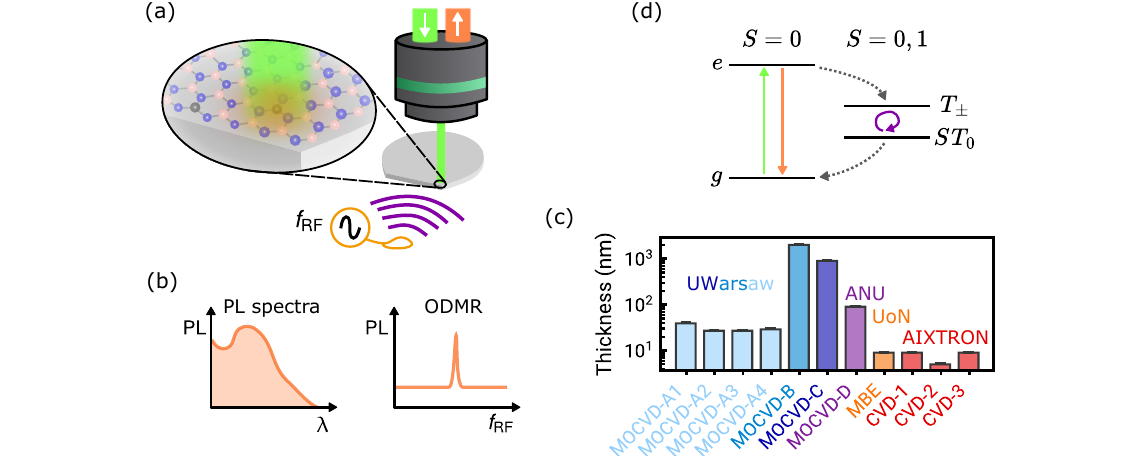}
\caption{\textbf{Concept and thickness of hBN samples.}
(a)~Schematic of measurement setup: PL from hBN deposited on a sapphire wafer is measured, irradiated with a 532\,nm laser and an RF field.
(b)~Schematic of ODMR and PL spectrum experiments undertaken to characterise samples. 
(c)~Measured thickness of surveyed hBN samples, with annotated sources (see Table~\ref{tab:1}), and estimated error of 5\% (see text and \SI).
(d)~Simplified Jablonski diagram, with excitation (green) and PL (orange) optical transitions, the driven spin pair transition (purple) in the metastable levels which are accessed via spin-dependent charge transfers (dotted).}
\label{fig:1}
\end{figure*}

Early work identified emitter densities increased with carbon doping during metal-organic chemical vapour deposition (MOCVD) or molecular beam epitaxy (MBE) growth, via conversion from graphite, and via carbon ion implantation~\cite{mendelsonIdentifyingCarbonSource2021}.
Later work demonstrated direct laser writing of spin pairs with deterministic placement~\cite{yangLaserDirectWriting2023}.
MOCVD studies have explored growth regimes favourable to different defect species (beyond just spin pairs), though without ODMR analysis that directly measures the optically active spin-half centres~\cite{dabrowskaDefectsLayeredBoron2024, iwanskiManipulatingCarbonRelated2024}.
More recently, carbon ion implantation has achieved deterministic creation of visible emitters with identical emission profiles, indicating specific classes of optical emitter origin~\cite{huaDeterministicCreationIdentical2024}: one tentatively identified as a carbon tetramer, the other with donor-acceptor pairs.
Donor-acceptor pairs are thought to be the origin of the spin pair emission, with the spin-active manifold a metastable state activated through charge-transfer~\cite{robertsonUniversalMechanismOptically2024}.
An oxygen-annealing process has been reported as an efficient generator of spin-active emitters with a high fraction displaying spin-one character~\cite{whitefieldGenerationNarrowbandQuantum2025}.
Finally, optical emitters in the visible range have been created electrically, and measured through electroluminescence~\cite{zhigulinElectricalGenerationColour2025}. 
Despite this broad scope of defect ensemble generation methods, there has to-date been no systematic study of ensemble spin and optical properties across different fabrication processes.
Here we will focus on growth from vapour phase, favoured due to their fine control over the fabrication conditions, and the potential for engineering large-area thin sensors approaching a monolayer thickness [Fig.~\ref{fig:1}(a)] ~\cite{sternSpectrallyResolvedPhotodynamics2019}.
Here we will study thicker films (3-2000\,nm) to facilitate comparison without being conflated by possible interface effects.

We begin with a survey of deposited films that were available to us classed based on the institution and technique of their providence, omitting samples without a discernible ODMR contrast.
We characterise the PL and ODMR response [Fig.~\ref{fig:1}(b)] of each spin ensemble to quantify their dc sensitivity to magnetic fields, normalised by the layer thickness with an eye to future magnetic imaging applications where not only the quality and quantity of spin defects sets the utility, but also the sample-sensor standoff.
We then investigate one preparation technique (MOCVD) as a function of growth temperature, and compare with previously reported electron spin resonance (ESR) spectra which provides useful context for our results.
Finally, we take the most sensitive hBN film and measure its ensemble $g$-factor, calibrated against a proximal NV-diamond sensor.

\section{Methods}

\subsection{Film deposition and thickness characterisation}

\begin{table*}[ht!]
\begin{tabularx}{0.99\textwidth}{ c c c Y Y Y Y Y Y Y Y Y } \toprule
 \multirow{2}{*}{Sample Name} & \multirow{2}{*}{Source} &            & Temp.       & P      &  TEB  & NH$_3$ & V/III &     PL              &                              ODMR &             ODMR    & Thick. \\ 
                              &                         &            & (\degree C) & (mbar) & ($\frac{\upmu \mathrm{mol}}{\mathrm{min}}$) & ($\frac{\mathrm{mmol}}{\mathrm{min}}$) & ratio & ($\dagger$) &  $\mathcal{C}$ (\textperthousand) & $w$ (MHz) & (nm) \\ \midrule
 MOCVD-A1 &                                    UWarsaw  &            &        1260 &    400 &  0-8.5 &  0-54 &  6274 &                109  &                 \hphantom{-}3.2  &                  52 &   39 \\ 
 MOCVD-A2 &                                    UWarsaw  &            &        1265 &    400 &  0-8.5 & 0-107 & 12548 &                110  &                 \hphantom{-}1.0  &                  55 &   27 \\ \cmidrule(r){3-8}
 \multirow{2}{*}{MOCVD-A3}  & \multirow{2}{*}{UWarsaw}  & \po G1\dd  &        1310 &    800 &  19   &  180  &  9411 & \multirow{2}{*}{42} & \multirow{2}{*}{\hphantom{-}0.4} & \multirow{2}{*}{73} & \multirow{2}{*}{27} \\
                            &                           & \po G2\pdd &        1295 &    400 &  0-7.1 & 0-110 & 15476 &                     &                                  &                     &  \\ \cmidrule(r){3-8}
 \multirow{2}{*}{MOCVD-A4}  & \multirow{2}{*}{UWarsaw}  & \po G1\dd  &        1310 &    800 &  19   &  180  &  9411 & \multirow{2}{*}{51} & \multirow{2}{*}{\hphantom{-}1.7} & \multirow{2}{*}{68} &  \multirow{2}{*}{29} \\
                            &                           & \po G2\pdd &        1295 &    400 &  0-7.1 & 0-110 & 15476 &                     &                                  &                     &  \\ \cmidrule(r){3-8}
 MOCVD-B  &                                    UWarsaw  &            &        1300 &    600 &  66   &  180  &  2689 &                  2  &                            -0.9  &                  41 & 2000 \\
 MOCVD-C  &                                    UWarsaw  &            &         820 &    100 &  150  &   4.5  &    29 &                 31  &                 \hphantom{-}6.4  &                 109 &  900 \\
 MOCVD-D    &                                        ANU  &            &        1350 &     85 &  0-30 &  0-45 &  4500 &             383  &                 \hphantom{-}3.0  &                  49 &  90 \\ 
 MBE      &                                    UoN  &            &            1390    &     -   &      - &  -     &  -     &                1134  &                 \hphantom{-}0.4  &                  39 & 9 \\
 CVD-1    &                                    AIXTRON  &            &        1350 &    500 &    -  &   -   & 1\ddd &                 89  &                 \hphantom{-}0.5  &                  88 &  9 \\ 
 CVD-2    &                                    AIXTRON  &            &        1350 &    500 &    -  &   -   & 1\ddd &                 50  &                 \hphantom{-}0.1  &                  74 &  3 \\
 CVD-3    &                                    AIXTRON  &            &        1350 &    500 &    -  &   -   & 1\ddd &                 94  &                 \hphantom{-}0.3  &                  65 &  9 \\ \bottomrule
\end{tabularx}
\caption{\textbf{hBN samples surveyed in Sec.~\ref{sec:survey}.} 
The samples are sourced from the University of Warsaw (UWarsaw), the Australian National University (ANU), the University of Nottingham (UoN) and AIXTRON SE. 
The precursor flow values given as a range (e.g.\ 0-30) represent the off/on state in the flow modulation growth scheme.
The V/III ratio given includes a factor to correct for relative pulse durations in the FME scheme.
($\dagger$)~Average PL (see text) in units of Mcounts/s/nm. 
($\ddagger$)~MOCVD-A3/4 are grown in two distinct stages (see text). 
(\S)~The CVD samples are grown with a single precursor borazine (B$_3$N$_3$H$_6$) at 8.7\,$\upmu$mol/min. 
} 
\label{tab:1}
\end{table*}

The chosen samples for our survey are described in Table~\ref{tab:1} alongside key performance metrics, and their thicknesses presented in Fig.~\ref{fig:1}(c).
Each of the MOCVD samples are grown in similar AIXTRON reactors and precursors, with the choice of growth parameters governing the observed properties. 

The MOCVD-\{A-C\} samples were grown on 2-inch sapphire substrates using an AIXTRON CCS 3×2” system. 
Triethylboron [TEB or (C$_2$H$_5$)$_3$B] and ammonia (NH$_3$) served as boron and nitrogen precursors, respectively, with hydrogen as the carrier gas. 
This study examines samples synthesised under different growth conditions. 
Samples MOCVD-A1 and MOCVD-A2 were fabricated using the flow modulation epitaxy (FME) method, in which precursors are introduced alternately into the reactor chamber~\cite{chughFlowModulationEpitaxy2018}. 
In contrast, samples MOCVD-A3 and MOCVD-A4 followed a two-stage epitaxial growth protocol described by Dąbrowska et al.~\cite{dabrowskaTwoStageEpitaxial2020}.
The first stage (G1) involved a 60-minute continuous flow growth (CFG) phase, where both precursors were supplied simultaneously~\cite{pakulaFundamentalMechanismsHBN2019}. 
G1 was followed by a second stage (G2), utilising the FME approach, similar to MOCVD-A1/A2.
Additionally, MOCVD-A4 underwent a 60-minute post-growth annealing in a nitrogen atmosphere. 
All four MOCVD-A samples formed planar layers oriented parallel to the sapphire substrate, exhibiting a characteristic wrinkle pattern resulting from difference in thermal expansion coefficient of BN and sapphire during cooling~\cite{iwanskiTemperatureInducedGiant2022, tatarczakStrainModulationEpitaxial2024}. 
Samples MOCVD-B and MOCVD-C were synthesised using the CFG regime, producing thick, porous structures~\cite{pakulaFundamentalMechanismsHBN2019, ciesielskiAllBNDistributedBragg2023}. 
MOCVD-C corresponds to sample S$_{820}^{\rm A}$ reported by Iwański et al.~\cite{iwanskiManipulatingCarbonRelated2024} and further studied in Sec.~\ref{sec:temps}.
The MOCVD-A sample thicknesses were measured by Fourier-transform infrared spectroscopy as decribed in \textcite{iwanskiTemperatureInducedGiant2022}.
The MOCVD-\{B,C\} sample thicknesses were estimated by in-situ optical reflectance during the growth process (see \textcite{ciesielskiAllBNDistributedBragg2023}).
The optical thickness measurements were confirmed by atomic force microscopy (AFM; see \SI).

The MOCVD-D sample corresponds to the `TEB-30' sample in Mendelson et al.~\cite{mendelsonIdentifyingCarbonSource2021}; it is grown on sapphire via an FME approach with peak TEB flow of 30\,$\upmu$mol/min, ammonia flow of 45\,mmol/min, in a hydrogen carrier gas at 1350\degree C.
The MOCVD-D sample thickness was characterised by AFM as 90\,nm thick~\cite{mendelsonIdentifyingCarbonSource2021}.

The MBE hBN layer was grown on sapphire by high-temperature molecular beam epitaxy (MBE) using a custom-designed dual chamber Veeco GENxplor MBE system. 
Details of the MBE growth procedure have been published previously~\cite{mendelsonIdentifyingCarbonSource2021, chengHightemperatureMolecularBeam2018}. 
We rely on thermocouple readings to measure the growth temperature of the substrate. 
For the sample discussed in the current paper the growth temperature was 1390\degree C. 
We used a high-temperature Knudsen effusion cell (Veeco) for the evaporation of isotopically purified $^{10}$B and a conventional Veeco RF plasma source to provide the active nitrogen flux. 
The carbon doping was achieved by electron beam evaporation of pyrolytic carbon using a Dr Eberl MBE-Komponenten source.
The hBN epilayers were grown using a nitrogen flow rate of 2\,sccm and a fixed RF power of 550\,W. 
We used $10\times10$\,mm$^2$ (0001) sapphire substrates. 
The thickness of the epilayer was determined \textit{ex-situ} by x-ray photoelectron spectroscopy (XPS) and variable-angle spectroscopic ellipsometry (VASE). 

The CVD samples were grown on c-plane sapphire (with a 0.2\degree\ off-cut toward the m-plane) in an AIXTRON CCS 2D R\&D reactor in a 19x2” configuration. 
A continuous hBN film was formed via a one-step growth process at 1350\degree C and 500\,mbar pressure using borazine (B$_3$N$_3$H$_6$; 8.7\,$\upmu$mol/min) as a precursor material. 
Sample CVD-1 was grown in H$_2$ atmosphere for 1600\,s, while samples CVD-2 and CVD-3 were grown using N$_2$ carrier gas for 140\,s and 280\,s, respectively.
The film thickness was measured by X-ray reflectivity, with the 3\,nm thickness of CVD-2 confirmed by transmission electron microscopy.

We normalise the surveyed samples' PL emission rate and sensitivity against sensor volume using the optical thickness measurements above.
To verify these optical measurements we etch a region of the hBN films and take an atomic-force microscope (AFM) image across the etch edge (see \SI).
We find broad agreement between optical and AFM thicknesses, with some discrepancies ascribed to porosity variations.
We conservatively assume a 5\% error on the optical thickness used for normalisation, from the distribution of AFM thicknesses measured on a single sample.

In Sec.\,\ref{sec:temps} we measure a series of MOCVD samples analogous to MOCVD-C, except with varying growth temperature.
This series of samples has been previously reported in Iwański et al.~\cite{iwanskiManipulatingCarbonRelated2024}, with extensive description and characterisation contained therein.
We note that the film grown at 820\degree C is six times thicker than the others in the series (due to longer growth time), which are all approximately 150\,nm thick.
For each growth temperature we measure two samples: one as-grown, another post-growth annealed in N$_2$ at 1200\degree C.

\subsection{PL Spectra}

To assess the optical properties of the spin ensembles we use a home-built widefield optical microscope~[Fig.~\ref{fig:1}(a)], equipped with an Ocean Insight Maya2000 Pro spectrometer, using typical illumination conditions for widefield sensing (532\,nm, peak power density 0.1\,mW\,$\upmu$m$^{-2}$).
The PL curves are shown with the sapphire substrate line masked, and normalised to their own maxima; they only express qualitative differences (see \SI\ for full raw dataset).

\subsection{Optically detected magnetic resonance spectra}

Spin measurements for the survey and temperature study were taken on the same purpose-built room temperature widefield microscopy setup [Fig.~\ref{fig:1}(a)]. 
A 532\,nm laser (Quantum Opus) provides spin initialisation and readout, gated by an acousto-optic modulator (Gooch \& Housego R35085-5). 
PL is collected through a microscope air objective (Nikon $20\times$ S Plan Fluor ELWD, $\mathrm{NA}=0.45$) and onto an sCMOS camera (Andor Zyla) via a longpass filter (550\,nm). 
RF radiation to manipulate the spin states is supplied by a Windfreak SynthNV PRO signal generator and amplified (Mini-Circuits HPA-50 W-63+). 
Camera exposures, laser pulses, and RF pulses are all sequenced using a SpinCore Pulseblaster ESR 500\,MHz card.
For all measurements, the samples were placed on a silver stripline (0.6\,mm wide) in a fixed position to ensure consistent RF power through the series.
A bias magnetic field of $\approx100$\,mT was aligned to the perpindicular to the substrate surface.
A continuous wave (CW) ODMR sequence was employed with an input 4\,W RF power. 
For all measurements we utilised an input laser power of 500\,mW, focused to a spot size of 25\,$\upmu$m diameter, for a peak laser power density of 0.1\,mW\,$\upmu$m$^{-2}$.
ODMR was acquired from this region (25\,$\upmu$m$\times$25$\upmu$m), where the PL intensity was maximal.

For all ODMR measurements we fit with a simple Lorentzian function, and calculated fit uncertainties with a bootstrap method~\cite{Bonamenta2017}, as follows.
The data is initially fit with a least squares method, then the standard deviation of the residuals is used as a measure of the noise in the measurement.
We then proceed to calculate random noise vectors and add these to the original measurement data and re-fit, 1000 times per ODMR. 
Fit parameter uncertainties are calculated from the standard deviation of the fit results for this set of 1000 samples.
We estimate a 10\% systematic uncertainty on the ODMR contrast measured due to optical bleaching of the spin ensemble; ODMR measurement integrations were kept short (a few minutes) to mitigate these effects.
These uncertainties are propagated through alongside the thickness uncertainty.

\subsection{$g$-factor measurement}

To measure the $g$-factor of the spin pairs [Fig.~\ref{fig:1}(d)] we choose MOCVD-D, the sample with the best dc magnetic sensitivity.
We also measure a micropowder sample, sourced from Graphene Supermarket, irradiated with 2\,MeV electrons and previously characterised in Scholten et al.\,\cite{scholtenMultispeciesOpticallyAddressable2023} `batch 2', as an isotropic equivalent.
The reference diamond is a $\langle 111 \rangle$ cut purchased from Element 6, implanted with $3\times10^{11}$cm$^{-2}$ Sb at 4\,MeV to produce a 1\,$\upmu$m vacancy layer, annealed under a ramp to 900\,\degree C, and acid boiled.

ODMR measurements for the $g$-factor measurement were taken in a similar setup to that described above except that parts of the setup (sample, circuit board, objective lens) were enclosed in a closed-cycle cryostat, as described in Refs.~\cite{lillieLaserModulationSuperconductivity2020, broadwayImagingDomainReversal2020}. 
The measurements were taken at the base temperature ($T=5$\,K) in a helium exchange gas.
The magnetic field was applied using a superconducting vector magnet, allowing us to precisely control both the direction and magnitude of the applied magnetic field.
This process allowed better than 1\degree\ alignment to the spin axis of an NV-diamond sample used for calibration.
RF radiation was supplied by an Agilent N5183A MXG signal generator and amplified by a Mini-Circuits ZVE-3W-183+, at a uniform power of 32\,W.
The input laser power into the cryostat was 180\,mW, and a 550-700\,nm bandpass filter was used on the collection arm.
Under these conditions a low-field two-peak B$_\mathrm{NV}$ measurement was taken to measure the zero-field splitting parameter $D$ of the NV spins, which varies with temperature, so that single resonance measurements could be taken at high-field.
To maintain as uniform conditions as possible, the NV-diamond, powder and MOCVD-D hBN measurements were taken at the same field without changing anything except the piezo position, to focus on each sequential sample.

\begin{figure*}[ht!]
\centering
\includegraphics[width=0.99\textwidth]{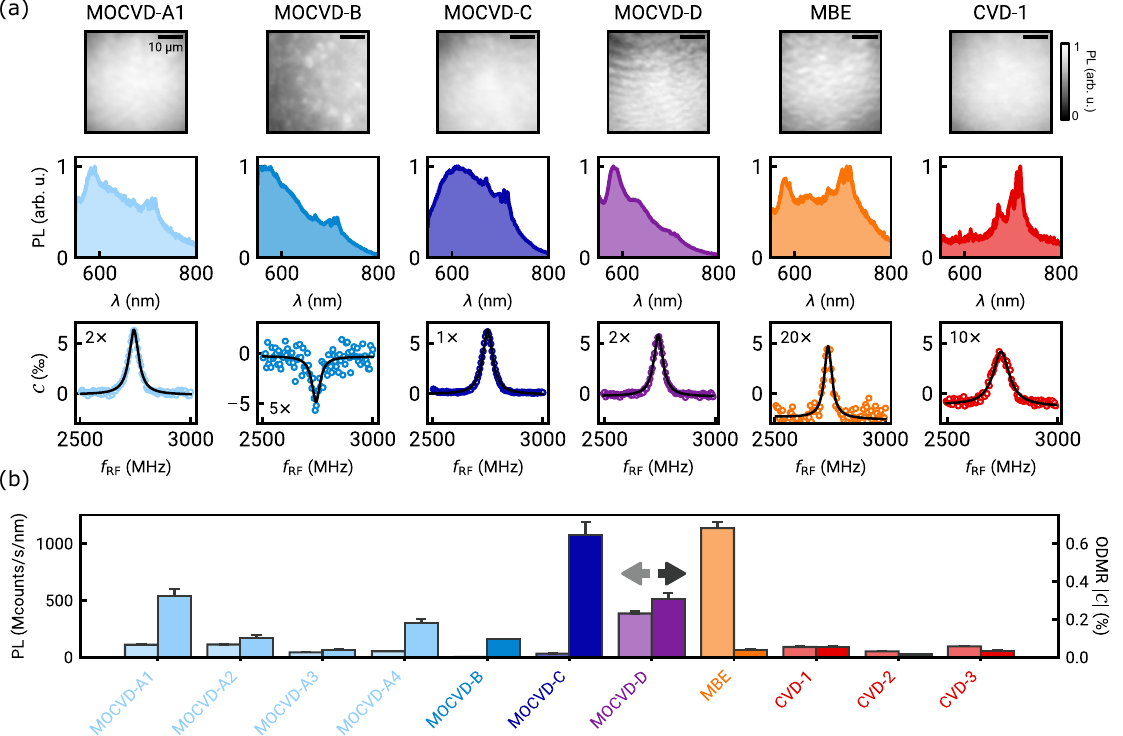}
\caption{\textbf{Survey of available spin-active hBN samples.}
(a)~PL image (top) spectra (middle) and ODMR spectra (bottom) of six exemplary samples. 
The ODMR contrast is non-zero off-resonance due to a cyclotron resonance background~\cite{wenOpticalSpinReadout2025, romestainOpticalDetectionCyclotron1980}.
(b)~PL normalised to sample thickness (left bar) and ODMR contrast (right bar) of all samples that displayed ODMR.
PL errors are from film thickness uncertainty, contrast errors from a statistical fit error (see text) and an additional 10\% to account for observed time-dependence of the ODMR signal (i.e.\ bleaching).}
\label{fig:2}
\end{figure*}

\section{Results}
\subsection{Survey of available deposited films}\label{sec:survey}

An example PL image, PL spectra and ODMR spectra from each sample class is shown in Fig.~\ref{fig:2}(a).
The PL emission is largely homogeneous across each sample and a representative region is selected for the following spectral analysis. 
The diverse range of emission structures between samples, and in particular the broad emission range from 550-800\,nm is characteristic of the emitter associated with spin pair ODMR, though we cannot discern anything useful from the features at this point.
The ODMR spectra display a single ODMR peak at the spin-half resonance $f_r = \gamma B_0$ for $\gamma \approx 28$\,GHz/T and a bias field strength $B_0 \approx100$\,mT.
The surveyed samples display a positive contrast 0.01-0.64\%, with the exception of MOCVD-B which is an outlier in samples we have measured to date as the sole bulk ensemble to give negative contrast ($-0.09\%$; also seen in micropowders~\cite{scholtenMultispeciesOpticallyAddressable2023}). 
The measured PL emission rate (normalised to thickness) and ODMR contrast ($\mathcal{C}$) are compared in Fig.~\ref{fig:2}(b). 
The MOCVD-C sample records the highest ODMR contrast (0.64\%) albeit with low PL emission (2\,Mcounts/s/nm), conversely the MBE sample records the highest PL emission (1134\,Mcounts/s/nm) with low ODMR contrast (0.04\%).
ODMR contrast is detectable down to the thinnest 3\,nm CVD sample (0.01\%).
There are no clear trends between PL and $\mathcal{C}$, though with attention to the growth parameters more information can be gleaned.
The highest PL samples (MOCVD-A1, MOCVD-D and MBE) have similar PL peaks below 600\,nm, and have similar lower V/III ratios, indicating more incorporation of carbon defects.
The highest $\mathcal{C}$ samples (MOCVD-A1, MOCVD-C and MOCVD-D) each have strong emission in the 600-650\,nm region, although this is contradicted by the MBE outlier; this might be explained by identifying a non-spin-dependent component in the 700\,nm emission region where the MBE sample (and the CVD) emit relatively strongly.  
The annealing step for MOCVD-A4 improved significantly on the ODMR contrast and PL emission compared to MOCVD-A3 (as-grown).
The significantly different growth regime (lower V/III ratio) of the MOCVD-B sample forms as misoriented flakes that are highly porous, which may explain the ensemble negative ODMR contrast which we have only measured in micropowders previously~\cite{scholtenMultispeciesOpticallyAddressable2023}, as well as reducing the PL emission per thickness.

\begin{figure}[t!]
\centering
\includegraphics[width=0.45\textwidth]{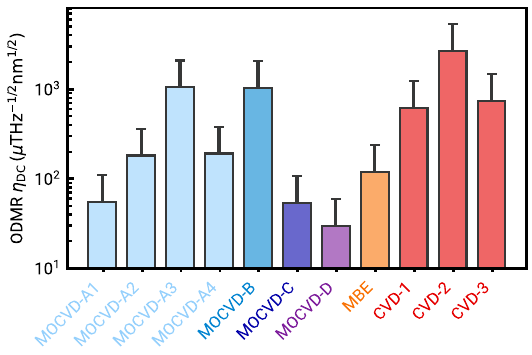}
\caption{\textbf{dc magnetic sensitivity of surveyed samples.}
Each measurement is normalised to the sample thickness.}
\label{fig:3}
\end{figure}

We calculate the dc magnetic sensitivity (Fig.~\ref{fig:3}) from the standard equation
\begin{equation}\label{eq:sens}
    \eta_\mathrm{dc} = \frac{w}{\gamma \mathcal{C} \sqrt{I_\mathrm{PL}}}
\end{equation}
where $w$ is the measured full-width at half-maximum resonance linewidth and $I_\mathrm{PL}$ is the PL emission rate per nanometre of thickness summed over a $25\,\upmu{\rm m} \times 25\,\upmu{\rm m}$ area~\cite{healeyComparisonDifferentMethods2020}. 
The linewidth (not shown) displays almost no variation, indicating that sensitivity variations are dominated by PL and $\mathcal{C}$.
The measured PL is dependent on emitter concentration and spin-pair displacement~\cite{robertsonUniversalMechanismOptically2024}.
The spin contrast is limited by any background emission without spin contribution, which is impossible to control in this survey, as well as the pair creation and recombination rates.
Further work may be able to correlate contrast with the PL emission features, which might be a route to further optimisation.
The MOCVD-D sample was measured as the lowest (best) sensitivity, combining strong PL emission (second best) with good ODMR contrast (second best).
The comparison of the MOCVD-C, MOCVD-D and MBE samples highlight that it is not just defect concentration (scaling with PL emission) that improves sensor performance, but the form of defect generation.
Future work is required to elucidate what growth parameters can improve ODMR contrast at high defect concentrations. 

\begin{figure*}[ht!]
\centering
\includegraphics[width=0.99\textwidth]{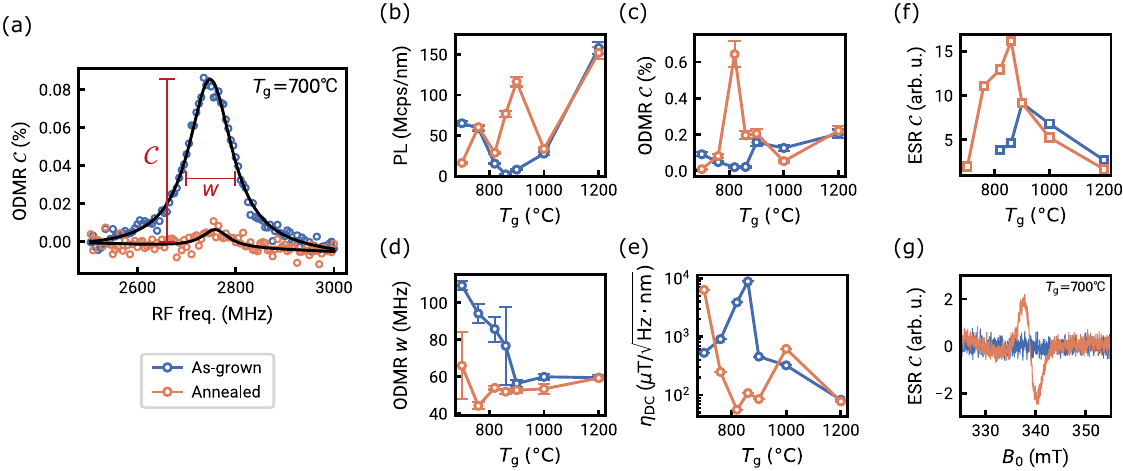}
\caption{\textbf{Characterising spin defects in MOCVD films as a function of growth temperature.}
(a)~ODMR for MOCVD film grown at 700\degree C (blue) and after common 1200\degree C N$_2$ annealing (peach). 
(b-e)~Properties of films as a function of growth temperature without (blue) and with (peach) annealing at 1200\degree C: PL (b), ODMR contrast (c) and width (d) as annotated in (a), and dc magnetic sensitivity (e) calculated according to Eq.~\ref{eq:sens}. 
(f)~ESR contrast for the same films as (b-e) reproduced from Iwański et al.~\cite{iwanskiManipulatingCarbonRelated2024}, showing a broad correlation with the ODMR results in (c). 
(g)~Example ESR signal for same samples as (a), in this case showing an ESR signal anticorrelated with the ODMR measurement.}
\label{fig:4}
\end{figure*}

\subsection{Characterising spin ensembles as a function of growth temperature}\label{sec:temps}

Having surveyed different deposition techniques for their spin ensemble properties, we now explore a singular technique (MOCVD) as a function of growth temperature.
We note that compared with the other techniques studied above, this series is grown in a significantly different regime -- in particular a low pressure (100\,mbar c.f.\ >400\,mbar) and very low V/III ratio ($\approx\,10$ c.f.\ $\approx\,1000$). 
This sample series is grown under 700-1200\degree C, with/without an additional post-growth annealing step at 1200\degree C in N$_2$ (see Methods).
These series' defect spin and optical properties have previously been studied via conventional ESR by Iwański et al.~\cite{iwanskiManipulatingCarbonRelated2024}, finding an optimal growth temperature of 860\degree C, improved by the annealing step.
This optimal temperature range was ascribed to the beginnings of crystal hBN structure (sp$^2$), with lower temperatures forming amorphous BC:N mixtures, and higher temperatures leading to improved BN crystal structure that limits defect creation.
Higher growth temperatures were also determined to redistribute hydrogens (via NH$_3$ pyrolysis) to the carbon clusters, consuming their free electrons~\cite{iwanskiManipulatingCarbonRelated2024}.
The annealing step is hypothesised to improve the crystal quality and thus the spin coherence, whilst retaining point defects and inclusions from the lower growth temperature.
However, the study by Iwański et al.~\cite{iwanskiManipulatingCarbonRelated2024} only measured the spin properties via bulk ESR, which cannot discriminate between the optically-addressable spin pairs and other $S=\sfrac{1}{2}$ defects.

ODMR [Fig.~\ref{fig:4}(a)] measurements allow us to directly measure the optically-active ensemble.
PL measurements across the series' [Fig.~\ref{fig:4}(b)] show a concave-up dependence on growth temperature, revived in the central temperature region by annealing. 
The large PL at the highest growth temperature can be explained as increased non-carbon defect numbers, in fact these samples display qualitatively different PL spectra (see Iwański et al.~\cite{iwanskiManipulatingCarbonRelated2024}).
ODMR contrasts [Fig.~\ref{fig:4}(c)] are relatively uniform, with a moderate maximum in the centre-temperature range again, though we note the 820\degree C sample is six times thicker than the rest of the series, potentially explaining the outlier result despite the thickness-normalisation, e.g.\ due to suppression of sapphire substrate PL or thickness-dependent PL collection~\cite{nishimuraInvestigationsOpticalAberration2024, scholtenAberrationControlQuantitative2022}.
Together, these results indicate defect creation decreases with growth temperature (except for an outlier 1200\degree growth), with annealing possibly helping to promote optically active defect formation in the central temperature range.
Increasing growth temperature reduces the ODMR linewidth for the unannealed series [Fig.~\ref{fig:4}(d)] down to a minimum width at 900\degree C.
We connect this improved linewidth to enhanced crystal quality reducing quasistatic magnetic noise, further supported by the annealed samples' consistently low linewidths.
The spin-half nature of the optically active defects means they are always on resonance with the spin-half bath, which may limit the ODMR contrast at low growth temperatures (with associated low crystal quality implying more free electrons) because cross-relaxation competes with optical polarisation [Fig.~\ref{fig:4}(c)].
To directly compare samples for sensing utility, we compare their dc magnetic sensitivity [Fig.~\ref{fig:4}(e)], with a clear region of optimality in the 800-900\degree C growth temperature range.

It is interesting to compare the ODMR results (which probe optically created metastable spin pairs) with the previous ESR results (which probe ground state $S=\sfrac{1}{2}$ spins).
Figure~\ref{fig:4}(f) shows a decreasing population of $S=\sfrac{1}{2}$ spins with growth temperature, with the ESR contrast reduced at low growth temperatures due to magnetic noise [see Fig.~\ref{fig:4}(d)].
Finally, to confirm the ensembles probed in ODMR and ESR are different, we show the ESR spectrum on the $T_{\rm g} = 700$\degree C sample measured via ODMR in Fig.~\ref{fig:4}(a) in Fig.~\ref{fig:4}(g).
Note that the as-grown sample displays significant ODMR contrast with minimal ESR contrast but the converse for after annealing, thus  indicating an anti-correlation between the two (ODMR, ESR) spin ensembles in this sample.

\begin{figure}[h!]
\centering
\includegraphics[width=0.4\textwidth]{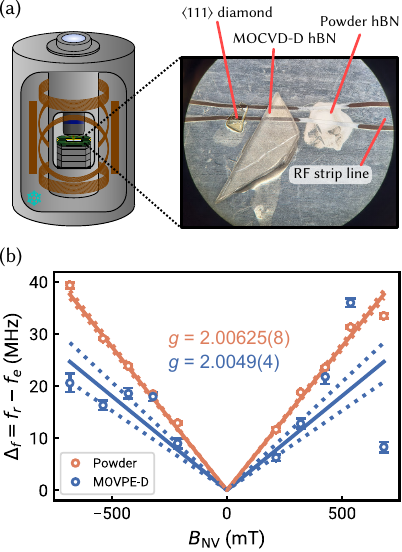}
\caption{\textbf{In-situ calibrated measurement of spin pair $g$-factor up to a half Tesla.}
(a)~A $\langle111\rangle$ NV-diamond, the MOCVD-D sample and a spin-active hBN powder are placed into a cryogenic microscope (5\,K) with a 1\,T vector magnet.
(b)~The difference ($\Delta_f$) between the measured hBN ODMR resonant frequency ($f_r$) and the free-electron $g$-factor ($f_e = 2.0023$) as a function of the bias field, measured at the NVs ($B_\mathrm{NV}$). 
Errors in B$_\mathrm{NV}$ (not observable on this scale) are due to NV alignment uncertainty ($<$1\degree), and statistical fit errors. 
Errors in $\Delta_f$ are estimated from statistical fit errors on the hBN ODMR spectrum fit (bootstrapped: see text).}
\label{fig:5}
\end{figure}

\subsection{Measurement of spin pair $g$-factor}\label{sec:gfac}

Though the previous ODMR spectra showed a spin-half behaviour with $g=2$, a more precise measurement of $g$ is lacking.
Under a magnetic field strength $B$, the ODMR resonance frequency ($f_r$) for a spin pair will be
\begin{equation}\label{eq:sp}
    f_r = \frac{g\,\mu_{\rm B}}{2\, h} B~,
\end{equation}
where $\mu_{\rm B}$ is the Bohr magneton, $h$ Planck's constant, and $g \approx 2$.
In particular, the free electron (one-electron cyclotron) $g$-factor is well known $g_e = 2.0023\dots$~\cite{fanMeasurementElectronMagnetic2023}.
Orbital contributions as well as higher order terms such as electronic spin-orbit interactions can shift the $g$-factor, or give it a non-isotropic character~\cite{dohertyTheoryGroundstateSpin2012}.
Here, we map ODMR spectra on our most sensitive deposited film (`MOCVD-D') and for comparison, a randomly oriented hBN powder film, as a function of applied field (210-680\,mT) to measure the $g$-factor.
We use an NV-diamond sample on the same sample holder [Fig.~\ref{fig:5}(a)] as a local highly accurate calibration of the bias field (see Methods).

The extracted resonant frequencies are shown in Fig.~\ref{fig:5}(b) as a difference from the free electron frequency, and fit to Eq.~\ref{eq:sp}.
Each series shows a clear positive gradient, indicating $g$-factors larger than the free electron, although with an unidentified source of error as some points lie outside the lines of worst fit.
We measure values of 2.0049(4) for the MOCVD-D sample and 2.00625(8) for the powder sample.
The consistent positive $\Delta_f = f_r - f_e$, where $f_e$ is the free-electron resonance ($g = g_e$ in Eq.~\ref{eq:sp}) indicates a small but measurable orbital component to the spin-pair $g$-factor, $\Delta_g = g - g_e > 2.6\times10^{-3}$.
We note that conventional ESR results found $g=2.0030$~\cite{iwanskiManipulatingCarbonRelated2024}, further highlighting that ESR and ODMR sample different spin systems. 
These preliminary results will guide future work to explore the gyromagnetic response of optically active hBN spin pairs.

\section{Conclusion}

In this work we have measured a range of hBN films deposited from vapour phase by scalable techniques for theirx optical spin properties, and found that they all displayed an ODMR response without irradiation or implantation. 
The survey of deposition techniques found that lower V/III ratios increased spin defect generation, which will guide future growth optimisation studies.
We found that the MOCVD-D sample used in many of the initial spin-pair studies~\cite{mendelsonIdentifyingCarbonSource2021, sternRoomtemperatureOpticallyDetected2022} remains the most magnetically sensitive, largely due to its high PL emission (defect concentration).
This best measured volume-normalised magnetic dc sensitivity of 30\,$\upmu$T\,Hz$^{-1/2}\upmu$m$^{3/2}$ is three orders of magnitude above that of a similarly thick $\delta$-doped NV layer at 25\,nT\,Hz$^{-1/2}\upmu$m$^{3/2}$~\cite{healeyComparisonDifferentMethods2020}, highlighting the need for further ensemble growth optimisation to be competitive in sensing applications. 

The ODMR analysis of the MOCVD growth temperature series found an optimal growth regime $\approx 820\degree$C, corroborating previous ESR measurements.
We showed that combining a moderate growth temperature with a subsequent high temperature annealing step to some extent combines the advantages of high point defect incorporation at low temperatures and promoting good crystallinity at high temperatures.
Future work could further optimise the growth process as a function of other parameters, for example by reducing the NH$_3$ supply in the flow modulation scheme as in the MOCVD-D sample, or post-growth annealing.
Finally, we measure the spin-pair $g$-factor against an NV-diamond ensemble up to a half Tesla, finding a consistent positive shift ($\Delta_g > 2.6\times10^{-3}$) from the free electron value.
This work provides a basis for further studies to understand the generation process of optical spin-pairs in hBN, and optimise their dc magnetic sensitivity.

\textbf{Acknowledgments}
The authors thank Alexander\,J.\,Healey for helpful discussions.
This work was supported by the Australian Research Council (ARC) through grants FT200100073, DP250100973 and CE200100010.
This work was performed in part at the Melbourne Centre for Nanofabrication (MCN) in the Victorian Node of the Australian National Fabrication Facility (ANFF) with M.S.F. ANFF-VIC Technology Fellowship.
Access to the MOCVD-D sample is through the ANFF ACT Node.
The work at Nottingham was supported by the Engineering and Physical Sciences Research Council UK (Grants No.\ EP/V05323X/1 and No.\ EP/W035510/1).
We acknowledge support from the grant of National Science Centre, Poland under decision 2022/47/B/ST5/03314.

\bibliography{refs}

\begin{thebibliography}{54}%
\makeatletter
\providecommand \@ifxundefined [1]{%
 \@ifx{#1\undefined}
}%
\providecommand \@ifnum [1]{%
 \ifnum #1\expandafter \@firstoftwo
 \else \expandafter \@secondoftwo
 \fi
}%
\providecommand \@ifx [1]{%
 \ifx #1\expandafter \@firstoftwo
 \else \expandafter \@secondoftwo
 \fi
}%
\providecommand \natexlab [1]{#1}%
\providecommand \enquote  [1]{``#1''}%
\providecommand \bibnamefont  [1]{#1}%
\providecommand \bibfnamefont [1]{#1}%
\providecommand \citenamefont [1]{#1}%
\providecommand \href@noop [0]{\@secondoftwo}%
\providecommand \href [0]{\begingroup \@sanitize@url \@href}%
\providecommand \@href[1]{\@@startlink{#1}\@@href}%
\providecommand \@@href[1]{\endgroup#1\@@endlink}%
\providecommand \@sanitize@url [0]{\catcode `\\12\catcode `\$12\catcode `\&12\catcode `\#12\catcode `\^12\catcode `\_12\catcode `\%12\relax}%
\providecommand \@@startlink[1]{}%
\providecommand \@@endlink[0]{}%
\providecommand \url  [0]{\begingroup\@sanitize@url \@url }%
\providecommand \@url [1]{\endgroup\@href {#1}{\urlprefix }}%
\providecommand \urlprefix  [0]{URL }%
\providecommand \Eprint [0]{\href }%
\providecommand \doibase [0]{https://doi.org/}%
\providecommand \selectlanguage [0]{\@gobble}%
\providecommand \bibinfo  [0]{\@secondoftwo}%
\providecommand \bibfield  [0]{\@secondoftwo}%
\providecommand \translation [1]{[#1]}%
\providecommand \BibitemOpen [0]{}%
\providecommand \bibitemStop [0]{}%
\providecommand \bibitemNoStop [0]{.\EOS\space}%
\providecommand \EOS [0]{\spacefactor3000\relax}%
\providecommand \BibitemShut  [1]{\csname bibitem#1\endcsname}%
\let\auto@bib@innerbib\@empty
\bibitem [{\citenamefont {Tetienne}(2021)}]{tetienneQuantumSensorsGo2021}%
  \BibitemOpen
  \bibfield  {author} {\bibinfo {author} {\bibfnamefont {J.-P.}\ \bibnamefont {Tetienne}},\ }\bibfield  {title} {\bibinfo {title} {Quantum sensors go flat},\ }\href {https://doi.org/10.1038/s41567-021-01338-5} {\bibfield  {journal} {\bibinfo  {journal} {Nature Physics}\ ,\ \bibinfo {pages} {1}} (\bibinfo {year} {2021})}\BibitemShut {NoStop}%
\bibitem [{\citenamefont {Liu}\ \emph {et~al.}(2024)\citenamefont {Liu}, \citenamefont {Li}, \citenamefont {Guo}, \citenamefont {Zeng}, \citenamefont {Xie}, \citenamefont {Liu}, \citenamefont {Ma}, \citenamefont {Wu}, \citenamefont {Wang}, \citenamefont {Wang}, \citenamefont {Ren}, \citenamefont {Ao}, \citenamefont {Xu}, \citenamefont {Tang}, \citenamefont {Gali}, \citenamefont {Li},\ and\ \citenamefont {Guo}}]{liuExperimentalObservationSpin2024}%
  \BibitemOpen
  \bibfield  {author} {\bibinfo {author} {\bibfnamefont {W.}~\bibnamefont {Liu}}, \bibinfo {author} {\bibfnamefont {S.}~\bibnamefont {Li}}, \bibinfo {author} {\bibfnamefont {N.-J.}\ \bibnamefont {Guo}}, \bibinfo {author} {\bibfnamefont {X.-D.}\ \bibnamefont {Zeng}}, \bibinfo {author} {\bibfnamefont {L.-K.}\ \bibnamefont {Xie}}, \bibinfo {author} {\bibfnamefont {J.-Y.}\ \bibnamefont {Liu}}, \bibinfo {author} {\bibfnamefont {Y.-H.}\ \bibnamefont {Ma}}, \bibinfo {author} {\bibfnamefont {Y.-Q.}\ \bibnamefont {Wu}}, \bibinfo {author} {\bibfnamefont {Y.-T.}\ \bibnamefont {Wang}}, \bibinfo {author} {\bibfnamefont {Z.-A.}\ \bibnamefont {Wang}}, \bibinfo {author} {\bibfnamefont {J.-M.}\ \bibnamefont {Ren}}, \bibinfo {author} {\bibfnamefont {C.}~\bibnamefont {Ao}}, \bibinfo {author} {\bibfnamefont {J.-S.}\ \bibnamefont {Xu}}, \bibinfo {author} {\bibfnamefont {J.-S.}\ \bibnamefont {Tang}}, \bibinfo {author} {\bibfnamefont {A.}~\bibnamefont {Gali}}, \bibinfo {author} {\bibfnamefont {C.-F.}\ \bibnamefont {Li}},\ and\
  \bibinfo {author} {\bibfnamefont {G.-C.}\ \bibnamefont {Guo}},\ }\href {https://doi.org/10.48550/arXiv.2410.18892} {\bibinfo {title} {Experimental observation of spin defects in van der {{Waals}} material {{GeS}}\$\_2\$}} (\bibinfo {year} {2024}),\ \Eprint {https://arxiv.org/abs/2410.18892} {arXiv:2410.18892} \BibitemShut {NoStop}%
\bibitem [{\citenamefont {Aharonovich}\ \emph {et~al.}(2022)\citenamefont {Aharonovich}, \citenamefont {Tetienne},\ and\ \citenamefont {Toth}}]{aharonovichQuantumEmittersHexagonal2022}%
  \BibitemOpen
  \bibfield  {author} {\bibinfo {author} {\bibfnamefont {I.}~\bibnamefont {Aharonovich}}, \bibinfo {author} {\bibfnamefont {J.-P.}\ \bibnamefont {Tetienne}},\ and\ \bibinfo {author} {\bibfnamefont {M.}~\bibnamefont {Toth}},\ }\bibfield  {title} {\bibinfo {title} {Quantum {{Emitters}} in {{Hexagonal Boron Nitride}}},\ }\href {https://doi.org/10.1021/acs.nanolett.2c03743} {\bibfield  {journal} {\bibinfo  {journal} {Nano Letters}\ ,\ \bibinfo {pages} {acs.nanolett.2c03743}} (\bibinfo {year} {2022})}\BibitemShut {NoStop}%
\bibitem [{\citenamefont {Healey}\ \emph {et~al.}(2022)\citenamefont {Healey}, \citenamefont {Scholten}, \citenamefont {Yang}, \citenamefont {Scott}, \citenamefont {Abrahams}, \citenamefont {Robertson}, \citenamefont {Hou}, \citenamefont {Guo}, \citenamefont {Rahman}, \citenamefont {Lu}, \citenamefont {Kianinia}, \citenamefont {Aharonovich},\ and\ \citenamefont {Tetienne}}]{healeyQuantumMicroscopyVan2022}%
  \BibitemOpen
  \bibfield  {author} {\bibinfo {author} {\bibfnamefont {A.~J.}\ \bibnamefont {Healey}}, \bibinfo {author} {\bibfnamefont {S.~C.}\ \bibnamefont {Scholten}}, \bibinfo {author} {\bibfnamefont {T.}~\bibnamefont {Yang}}, \bibinfo {author} {\bibfnamefont {J.~A.}\ \bibnamefont {Scott}}, \bibinfo {author} {\bibfnamefont {G.~J.}\ \bibnamefont {Abrahams}}, \bibinfo {author} {\bibfnamefont {I.~O.}\ \bibnamefont {Robertson}}, \bibinfo {author} {\bibfnamefont {X.~F.}\ \bibnamefont {Hou}}, \bibinfo {author} {\bibfnamefont {Y.~F.}\ \bibnamefont {Guo}}, \bibinfo {author} {\bibfnamefont {S.}~\bibnamefont {Rahman}}, \bibinfo {author} {\bibfnamefont {Y.}~\bibnamefont {Lu}}, \bibinfo {author} {\bibfnamefont {M.}~\bibnamefont {Kianinia}}, \bibinfo {author} {\bibfnamefont {I.}~\bibnamefont {Aharonovich}},\ and\ \bibinfo {author} {\bibfnamefont {J.-P.}\ \bibnamefont {Tetienne}},\ }\bibfield  {title} {\bibinfo {title} {Quantum microscopy with van der {{Waals}} heterostructures},\ }\href {https://doi.org/10.1038/s41567-022-01815-5}
  {\bibfield  {journal} {\bibinfo  {journal} {Nature Physics}\ ,\ \bibinfo {pages} {1}} (\bibinfo {year} {2022})}\BibitemShut {NoStop}%
\bibitem [{\citenamefont {Ghiasi}\ \emph {et~al.}(2023)\citenamefont {Ghiasi}, \citenamefont {Borst}, \citenamefont {Kurdi}, \citenamefont {Simon}, \citenamefont {Bertelli}, \citenamefont {{Boix-Constant}}, \citenamefont {{Ma{\~n}as-Valero}}, \citenamefont {{van der Zant}},\ and\ \citenamefont {{van der Sar}}}]{ghiasiNitrogenvacancyMagnetometryCrSBr2023}%
  \BibitemOpen
  \bibfield  {author} {\bibinfo {author} {\bibfnamefont {T.~S.}\ \bibnamefont {Ghiasi}}, \bibinfo {author} {\bibfnamefont {M.}~\bibnamefont {Borst}}, \bibinfo {author} {\bibfnamefont {S.}~\bibnamefont {Kurdi}}, \bibinfo {author} {\bibfnamefont {B.~G.}\ \bibnamefont {Simon}}, \bibinfo {author} {\bibfnamefont {I.}~\bibnamefont {Bertelli}}, \bibinfo {author} {\bibfnamefont {C.}~\bibnamefont {{Boix-Constant}}}, \bibinfo {author} {\bibfnamefont {S.}~\bibnamefont {{Ma{\~n}as-Valero}}}, \bibinfo {author} {\bibfnamefont {H.~S.~J.}\ \bibnamefont {{van der Zant}}},\ and\ \bibinfo {author} {\bibfnamefont {T.}~\bibnamefont {{van der Sar}}},\ }\bibfield  {title} {\bibinfo {title} {Nitrogen-vacancy magnetometry of {{CrSBr}} by diamond membrane transfer},\ }\href {https://doi.org/10.1038/s41699-023-00423-y} {\bibfield  {journal} {\bibinfo  {journal} {npj 2D Materials and Applications}\ }\textbf {\bibinfo {volume} {7}},\ \bibinfo {pages} {1} (\bibinfo {year} {2023})}\BibitemShut {NoStop}%
\bibitem [{\citenamefont {Abrahams}\ \emph {et~al.}(2021)\citenamefont {Abrahams}, \citenamefont {Scholten}, \citenamefont {Healey}, \citenamefont {Robertson}, \citenamefont {Dontschuk}, \citenamefont {Lim}, \citenamefont {Johnson}, \citenamefont {Simpson}, \citenamefont {Hollenberg},\ and\ \citenamefont {Tetienne}}]{abrahamsIntegratedWidefieldProbe2021}%
  \BibitemOpen
  \bibfield  {author} {\bibinfo {author} {\bibfnamefont {G.~J.}\ \bibnamefont {Abrahams}}, \bibinfo {author} {\bibfnamefont {S.~C.}\ \bibnamefont {Scholten}}, \bibinfo {author} {\bibfnamefont {A.~J.}\ \bibnamefont {Healey}}, \bibinfo {author} {\bibfnamefont {I.~O.}\ \bibnamefont {Robertson}}, \bibinfo {author} {\bibfnamefont {N.}~\bibnamefont {Dontschuk}}, \bibinfo {author} {\bibfnamefont {S.~Q.}\ \bibnamefont {Lim}}, \bibinfo {author} {\bibfnamefont {B.~C.}\ \bibnamefont {Johnson}}, \bibinfo {author} {\bibfnamefont {D.~A.}\ \bibnamefont {Simpson}}, \bibinfo {author} {\bibfnamefont {L.~C.~L.}\ \bibnamefont {Hollenberg}},\ and\ \bibinfo {author} {\bibfnamefont {J.-P.}\ \bibnamefont {Tetienne}},\ }\bibfield  {title} {\bibinfo {title} {An integrated widefield probe for practical diamond nitrogen-vacancy microscopy},\ }\href {https://doi.org/10.1063/5.0073320} {\bibfield  {journal} {\bibinfo  {journal} {Applied Physics Letters}\ }\textbf {\bibinfo {volume} {119}},\ \bibinfo {pages} {254002} (\bibinfo {year}
  {2021})}\BibitemShut {NoStop}%
\bibitem [{\citenamefont {Yao}\ \emph {et~al.}(2018)\citenamefont {Yao}, \citenamefont {Zaletel}, \citenamefont {{Stamper-Kurn}},\ and\ \citenamefont {Vishwanath}}]{yaoQuantumDipolarSpin2018}%
  \BibitemOpen
  \bibfield  {author} {\bibinfo {author} {\bibfnamefont {N.~Y.}\ \bibnamefont {Yao}}, \bibinfo {author} {\bibfnamefont {M.~P.}\ \bibnamefont {Zaletel}}, \bibinfo {author} {\bibfnamefont {D.~M.}\ \bibnamefont {{Stamper-Kurn}}},\ and\ \bibinfo {author} {\bibfnamefont {A.}~\bibnamefont {Vishwanath}},\ }\bibfield  {title} {\bibinfo {title} {A quantum dipolar spin liquid},\ }\href {https://doi.org/10.1038/s41567-017-0030-7} {\bibfield  {journal} {\bibinfo  {journal} {Nature Physics}\ }\textbf {\bibinfo {volume} {14}},\ \bibinfo {pages} {405} (\bibinfo {year} {2018})}\BibitemShut {NoStop}%
\bibitem [{\citenamefont {Davis}\ \emph {et~al.}(2023)\citenamefont {Davis}, \citenamefont {Ye}, \citenamefont {Machado}, \citenamefont {Meynell}, \citenamefont {Wu}, \citenamefont {Mittiga}, \citenamefont {Schenken}, \citenamefont {Joos}, \citenamefont {Kobrin}, \citenamefont {Lyu}, \citenamefont {Wang}, \citenamefont {Bluvstein}, \citenamefont {Choi}, \citenamefont {Zu}, \citenamefont {Jayich},\ and\ \citenamefont {Yao}}]{davisProbingManybodyDynamics2023}%
  \BibitemOpen
  \bibfield  {author} {\bibinfo {author} {\bibfnamefont {E.~J.}\ \bibnamefont {Davis}}, \bibinfo {author} {\bibfnamefont {B.}~\bibnamefont {Ye}}, \bibinfo {author} {\bibfnamefont {F.}~\bibnamefont {Machado}}, \bibinfo {author} {\bibfnamefont {S.~A.}\ \bibnamefont {Meynell}}, \bibinfo {author} {\bibfnamefont {W.}~\bibnamefont {Wu}}, \bibinfo {author} {\bibfnamefont {T.}~\bibnamefont {Mittiga}}, \bibinfo {author} {\bibfnamefont {W.}~\bibnamefont {Schenken}}, \bibinfo {author} {\bibfnamefont {M.}~\bibnamefont {Joos}}, \bibinfo {author} {\bibfnamefont {B.}~\bibnamefont {Kobrin}}, \bibinfo {author} {\bibfnamefont {Y.}~\bibnamefont {Lyu}}, \bibinfo {author} {\bibfnamefont {Z.}~\bibnamefont {Wang}}, \bibinfo {author} {\bibfnamefont {D.}~\bibnamefont {Bluvstein}}, \bibinfo {author} {\bibfnamefont {S.}~\bibnamefont {Choi}}, \bibinfo {author} {\bibfnamefont {C.}~\bibnamefont {Zu}}, \bibinfo {author} {\bibfnamefont {A.~C.~B.}\ \bibnamefont {Jayich}},\ and\ \bibinfo {author} {\bibfnamefont {N.~Y.}\ \bibnamefont {Yao}},\
  }\bibfield  {title} {\bibinfo {title} {Probing many-body dynamics in a two-dimensional dipolar spin ensemble},\ }\href {https://doi.org/10.1038/s41567-023-01944-5} {\bibfield  {journal} {\bibinfo  {journal} {Nature Physics}\ }\textbf {\bibinfo {volume} {19}},\ \bibinfo {pages} {836} (\bibinfo {year} {2023})}\BibitemShut {NoStop}%
\bibitem [{\citenamefont {Abanin}\ \emph {et~al.}(2019)\citenamefont {Abanin}, \citenamefont {Altman}, \citenamefont {Bloch},\ and\ \citenamefont {Serbyn}}]{abaninColloquiumManybodyLocalization2019}%
  \BibitemOpen
  \bibfield  {author} {\bibinfo {author} {\bibfnamefont {D.~A.}\ \bibnamefont {Abanin}}, \bibinfo {author} {\bibfnamefont {E.}~\bibnamefont {Altman}}, \bibinfo {author} {\bibfnamefont {I.}~\bibnamefont {Bloch}},\ and\ \bibinfo {author} {\bibfnamefont {M.}~\bibnamefont {Serbyn}},\ }\bibfield  {title} {\bibinfo {title} {Colloquium: {{Many-body}} localization, thermalization, and entanglement},\ }\href {https://doi.org/10.1103/RevModPhys.91.021001} {\bibfield  {journal} {\bibinfo  {journal} {Reviews of Modern Physics}\ }\textbf {\bibinfo {volume} {91}},\ \bibinfo {pages} {021001} (\bibinfo {year} {2019})}\BibitemShut {NoStop}%
\bibitem [{\citenamefont {Gottscholl}\ \emph {et~al.}(2021)\citenamefont {Gottscholl}, \citenamefont {Diez}, \citenamefont {Soltamov}, \citenamefont {Kasper}, \citenamefont {Krau{\ss}e}, \citenamefont {Sperlich}, \citenamefont {Kianinia}, \citenamefont {Bradac}, \citenamefont {Aharonovich},\ and\ \citenamefont {Dyakonov}}]{gottschollSpinDefectsHBN2021}%
  \BibitemOpen
  \bibfield  {author} {\bibinfo {author} {\bibfnamefont {A.}~\bibnamefont {Gottscholl}}, \bibinfo {author} {\bibfnamefont {M.}~\bibnamefont {Diez}}, \bibinfo {author} {\bibfnamefont {V.}~\bibnamefont {Soltamov}}, \bibinfo {author} {\bibfnamefont {C.}~\bibnamefont {Kasper}}, \bibinfo {author} {\bibfnamefont {D.}~\bibnamefont {Krau{\ss}e}}, \bibinfo {author} {\bibfnamefont {A.}~\bibnamefont {Sperlich}}, \bibinfo {author} {\bibfnamefont {M.}~\bibnamefont {Kianinia}}, \bibinfo {author} {\bibfnamefont {C.}~\bibnamefont {Bradac}}, \bibinfo {author} {\bibfnamefont {I.}~\bibnamefont {Aharonovich}},\ and\ \bibinfo {author} {\bibfnamefont {V.}~\bibnamefont {Dyakonov}},\ }\bibfield  {title} {\bibinfo {title} {Spin defects in {{hBN}} as promising temperature, pressure and magnetic field quantum sensors},\ }\href {https://doi.org/10.1038/s41467-021-24725-1} {\bibfield  {journal} {\bibinfo  {journal} {Nature Communications}\ }\textbf {\bibinfo {volume} {12}},\ \bibinfo {pages} {4480} (\bibinfo {year} {2021})}\BibitemShut
  {NoStop}%
\bibitem [{\citenamefont {Huang}\ \emph {et~al.}(2022)\citenamefont {Huang}, \citenamefont {Zhou}, \citenamefont {Chen}, \citenamefont {Lu}, \citenamefont {McLaughlin}, \citenamefont {Li}, \citenamefont {Alghamdi}, \citenamefont {Djugba}, \citenamefont {Shi}, \citenamefont {Wang},\ and\ \citenamefont {Du}}]{huangWideFieldImaging2022}%
  \BibitemOpen
  \bibfield  {author} {\bibinfo {author} {\bibfnamefont {M.}~\bibnamefont {Huang}}, \bibinfo {author} {\bibfnamefont {J.}~\bibnamefont {Zhou}}, \bibinfo {author} {\bibfnamefont {D.}~\bibnamefont {Chen}}, \bibinfo {author} {\bibfnamefont {H.}~\bibnamefont {Lu}}, \bibinfo {author} {\bibfnamefont {N.~J.}\ \bibnamefont {McLaughlin}}, \bibinfo {author} {\bibfnamefont {S.}~\bibnamefont {Li}}, \bibinfo {author} {\bibfnamefont {M.}~\bibnamefont {Alghamdi}}, \bibinfo {author} {\bibfnamefont {D.}~\bibnamefont {Djugba}}, \bibinfo {author} {\bibfnamefont {J.}~\bibnamefont {Shi}}, \bibinfo {author} {\bibfnamefont {H.}~\bibnamefont {Wang}},\ and\ \bibinfo {author} {\bibfnamefont {C.~R.}\ \bibnamefont {Du}},\ }\bibfield  {title} {\bibinfo {title} {Wide field imaging of van der {{Waals}} ferromagnet {{Fe3GeTe2}} by spin defects in hexagonal boron nitride},\ }\href {https://doi.org/10.1038/s41467-022-33016-2} {\bibfield  {journal} {\bibinfo  {journal} {Nature Communications}\ }\textbf {\bibinfo {volume} {13}},\ \bibinfo
  {pages} {5369} (\bibinfo {year} {2022})}\BibitemShut {NoStop}%
\bibitem [{\citenamefont {Kumar}\ \emph {et~al.}(2022)\citenamefont {Kumar}, \citenamefont {Fabre}, \citenamefont {Durand}, \citenamefont {{Clua-Provost}}, \citenamefont {Li}, \citenamefont {Edgar}, \citenamefont {Rougemaille}, \citenamefont {Coraux}, \citenamefont {Marie}, \citenamefont {Renucci}, \citenamefont {Robert}, \citenamefont {{Robert-Philip}}, \citenamefont {Gil}, \citenamefont {Cassabois}, \citenamefont {Finco},\ and\ \citenamefont {Jacques}}]{kumarMagneticImagingSpin2022}%
  \BibitemOpen
  \bibfield  {author} {\bibinfo {author} {\bibfnamefont {P.}~\bibnamefont {Kumar}}, \bibinfo {author} {\bibfnamefont {F.}~\bibnamefont {Fabre}}, \bibinfo {author} {\bibfnamefont {A.}~\bibnamefont {Durand}}, \bibinfo {author} {\bibfnamefont {T.}~\bibnamefont {{Clua-Provost}}}, \bibinfo {author} {\bibfnamefont {J.}~\bibnamefont {Li}}, \bibinfo {author} {\bibfnamefont {J.}~\bibnamefont {Edgar}}, \bibinfo {author} {\bibfnamefont {N.}~\bibnamefont {Rougemaille}}, \bibinfo {author} {\bibfnamefont {J.}~\bibnamefont {Coraux}}, \bibinfo {author} {\bibfnamefont {X.}~\bibnamefont {Marie}}, \bibinfo {author} {\bibfnamefont {P.}~\bibnamefont {Renucci}}, \bibinfo {author} {\bibfnamefont {C.}~\bibnamefont {Robert}}, \bibinfo {author} {\bibfnamefont {I.}~\bibnamefont {{Robert-Philip}}}, \bibinfo {author} {\bibfnamefont {B.}~\bibnamefont {Gil}}, \bibinfo {author} {\bibfnamefont {G.}~\bibnamefont {Cassabois}}, \bibinfo {author} {\bibfnamefont {A.}~\bibnamefont {Finco}},\ and\ \bibinfo {author} {\bibfnamefont {V.}~\bibnamefont
  {Jacques}},\ }\bibfield  {title} {\bibinfo {title} {Magnetic {{Imaging}} with {{Spin Defects}} in {{Hexagonal Boron Nitride}}},\ }\href {https://doi.org/10.1103/PhysRevApplied.18.L061002} {\bibfield  {journal} {\bibinfo  {journal} {Physical Review Applied}\ }\textbf {\bibinfo {volume} {18}},\ \bibinfo {pages} {L061002} (\bibinfo {year} {2022})},\ \Eprint {https://arxiv.org/abs/2207.10477} {arXiv:2207.10477 [cond-mat]} \BibitemShut {NoStop}%
\bibitem [{\citenamefont {Gong}\ \emph {et~al.}(2022)\citenamefont {Gong}, \citenamefont {He}, \citenamefont {Gao}, \citenamefont {Ju}, \citenamefont {Liu}, \citenamefont {Ye}, \citenamefont {Henriksen}, \citenamefont {Li},\ and\ \citenamefont {Zu}}]{gongCoherentDynamicsStrongly2022}%
  \BibitemOpen
  \bibfield  {author} {\bibinfo {author} {\bibfnamefont {R.}~\bibnamefont {Gong}}, \bibinfo {author} {\bibfnamefont {G.}~\bibnamefont {He}}, \bibinfo {author} {\bibfnamefont {X.}~\bibnamefont {Gao}}, \bibinfo {author} {\bibfnamefont {P.}~\bibnamefont {Ju}}, \bibinfo {author} {\bibfnamefont {Z.}~\bibnamefont {Liu}}, \bibinfo {author} {\bibfnamefont {B.}~\bibnamefont {Ye}}, \bibinfo {author} {\bibfnamefont {E.~A.}\ \bibnamefont {Henriksen}}, \bibinfo {author} {\bibfnamefont {T.}~\bibnamefont {Li}},\ and\ \bibinfo {author} {\bibfnamefont {C.}~\bibnamefont {Zu}},\ }\href {https://doi.org/10.48550/arXiv.2210.11485} {\bibinfo {title} {Coherent {{Dynamics}} of {{Strongly Interacting Electronic Spin Defects}} in {{Hexagonal Boron Nitride}}}} (\bibinfo {year} {2022}),\ \Eprint {https://arxiv.org/abs/2210.11485} {arXiv:2210.11485 [cond-mat, physics:quant-ph]} \BibitemShut {NoStop}%
\bibitem [{\citenamefont {Curie}\ \emph {et~al.}(2022)\citenamefont {Curie}, \citenamefont {Krogel}, \citenamefont {Cavar}, \citenamefont {Solanki}, \citenamefont {Upadhyaya}, \citenamefont {Li}, \citenamefont {Pai}, \citenamefont {Chilcote}, \citenamefont {Iyer}, \citenamefont {Puretzky}, \citenamefont {Ivanov}, \citenamefont {Du}, \citenamefont {Reboredo},\ and\ \citenamefont {Lawrie}}]{curieCorrelativeNanoscaleImaging2022}%
  \BibitemOpen
  \bibfield  {author} {\bibinfo {author} {\bibfnamefont {D.}~\bibnamefont {Curie}}, \bibinfo {author} {\bibfnamefont {J.~T.}\ \bibnamefont {Krogel}}, \bibinfo {author} {\bibfnamefont {L.}~\bibnamefont {Cavar}}, \bibinfo {author} {\bibfnamefont {A.}~\bibnamefont {Solanki}}, \bibinfo {author} {\bibfnamefont {P.}~\bibnamefont {Upadhyaya}}, \bibinfo {author} {\bibfnamefont {T.}~\bibnamefont {Li}}, \bibinfo {author} {\bibfnamefont {Y.-Y.}\ \bibnamefont {Pai}}, \bibinfo {author} {\bibfnamefont {M.}~\bibnamefont {Chilcote}}, \bibinfo {author} {\bibfnamefont {V.}~\bibnamefont {Iyer}}, \bibinfo {author} {\bibfnamefont {A.}~\bibnamefont {Puretzky}}, \bibinfo {author} {\bibfnamefont {I.}~\bibnamefont {Ivanov}}, \bibinfo {author} {\bibfnamefont {M.-H.}\ \bibnamefont {Du}}, \bibinfo {author} {\bibfnamefont {F.}~\bibnamefont {Reboredo}},\ and\ \bibinfo {author} {\bibfnamefont {B.}~\bibnamefont {Lawrie}},\ }\href {http://arxiv.org/abs/2203.10075} {\bibinfo {title} {Correlative nanoscale imaging of strained {{hBN}} spin
  defects}} (\bibinfo {year} {2022}),\ \Eprint {https://arxiv.org/abs/2203.10075} {arXiv:2203.10075 [cond-mat, physics:quant-ph]} \BibitemShut {NoStop}%
\bibitem [{\citenamefont {Lyu}\ \emph {et~al.}(2022)\citenamefont {Lyu}, \citenamefont {Tan}, \citenamefont {Wu}, \citenamefont {Zhang}, \citenamefont {Zhang}, \citenamefont {Mu}, \citenamefont {{Z{\'u}{\~n}iga-P{\'e}rez}}, \citenamefont {Cai},\ and\ \citenamefont {Gao}}]{lyuStrainQuantumSensing2022}%
  \BibitemOpen
  \bibfield  {author} {\bibinfo {author} {\bibfnamefont {X.}~\bibnamefont {Lyu}}, \bibinfo {author} {\bibfnamefont {Q.}~\bibnamefont {Tan}}, \bibinfo {author} {\bibfnamefont {L.}~\bibnamefont {Wu}}, \bibinfo {author} {\bibfnamefont {C.}~\bibnamefont {Zhang}}, \bibinfo {author} {\bibfnamefont {Z.}~\bibnamefont {Zhang}}, \bibinfo {author} {\bibfnamefont {Z.}~\bibnamefont {Mu}}, \bibinfo {author} {\bibfnamefont {J.}~\bibnamefont {{Z{\'u}{\~n}iga-P{\'e}rez}}}, \bibinfo {author} {\bibfnamefont {H.}~\bibnamefont {Cai}},\ and\ \bibinfo {author} {\bibfnamefont {W.}~\bibnamefont {Gao}},\ }\bibfield  {title} {\bibinfo {title} {Strain {{Quantum Sensing}} with {{Spin Defects}} in {{Hexagonal Boron Nitride}}},\ }\href {https://doi.org/10.1021/acs.nanolett.2c01722} {\bibfield  {journal} {\bibinfo  {journal} {Nano Letters}\ ,\ \bibinfo {pages} {acs.nanolett.2c01722}} (\bibinfo {year} {2022})}\BibitemShut {NoStop}%
\bibitem [{\citenamefont {Yang}\ \emph {et~al.}(2022)\citenamefont {Yang}, \citenamefont {Mendelson}, \citenamefont {Li}, \citenamefont {Gottscholl}, \citenamefont {Scott}, \citenamefont {Kianinia}, \citenamefont {Dyakonov}, \citenamefont {Toth},\ and\ \citenamefont {Aharonovich}}]{yangSpinDefectsHexagonal2022}%
  \BibitemOpen
  \bibfield  {author} {\bibinfo {author} {\bibfnamefont {T.}~\bibnamefont {Yang}}, \bibinfo {author} {\bibfnamefont {N.}~\bibnamefont {Mendelson}}, \bibinfo {author} {\bibfnamefont {C.}~\bibnamefont {Li}}, \bibinfo {author} {\bibfnamefont {A.}~\bibnamefont {Gottscholl}}, \bibinfo {author} {\bibfnamefont {J.}~\bibnamefont {Scott}}, \bibinfo {author} {\bibfnamefont {M.}~\bibnamefont {Kianinia}}, \bibinfo {author} {\bibfnamefont {V.}~\bibnamefont {Dyakonov}}, \bibinfo {author} {\bibfnamefont {M.}~\bibnamefont {Toth}},\ and\ \bibinfo {author} {\bibfnamefont {I.}~\bibnamefont {Aharonovich}},\ }\bibfield  {title} {\bibinfo {title} {Spin defects in hexagonal boron nitride for strain sensing on nanopillar arrays},\ }\href {https://doi.org/10.1039/D1NR07919K} {\bibfield  {journal} {\bibinfo  {journal} {Nanoscale}\ }\textbf {\bibinfo {volume} {14}},\ \bibinfo {pages} {5239} (\bibinfo {year} {2022})}\BibitemShut {NoStop}%
\bibitem [{\citenamefont {Guo}\ \emph {et~al.}(2023)\citenamefont {Guo}, \citenamefont {Li}, \citenamefont {Liu}, \citenamefont {Yang}, \citenamefont {Zeng}, \citenamefont {Yu}, \citenamefont {Meng}, \citenamefont {Li}, \citenamefont {Wang}, \citenamefont {Xie}, \citenamefont {Ge}, \citenamefont {Wang}, \citenamefont {Li}, \citenamefont {Xu}, \citenamefont {Wang}, \citenamefont {Tang}, \citenamefont {Gali}, \citenamefont {Li},\ and\ \citenamefont {Guo}}]{guoCoherentControlUltrabright2023}%
  \BibitemOpen
  \bibfield  {author} {\bibinfo {author} {\bibfnamefont {N.-J.}\ \bibnamefont {Guo}}, \bibinfo {author} {\bibfnamefont {S.}~\bibnamefont {Li}}, \bibinfo {author} {\bibfnamefont {W.}~\bibnamefont {Liu}}, \bibinfo {author} {\bibfnamefont {Y.-Z.}\ \bibnamefont {Yang}}, \bibinfo {author} {\bibfnamefont {X.-D.}\ \bibnamefont {Zeng}}, \bibinfo {author} {\bibfnamefont {S.}~\bibnamefont {Yu}}, \bibinfo {author} {\bibfnamefont {Y.}~\bibnamefont {Meng}}, \bibinfo {author} {\bibfnamefont {Z.-P.}\ \bibnamefont {Li}}, \bibinfo {author} {\bibfnamefont {Z.-A.}\ \bibnamefont {Wang}}, \bibinfo {author} {\bibfnamefont {L.-K.}\ \bibnamefont {Xie}}, \bibinfo {author} {\bibfnamefont {R.-C.}\ \bibnamefont {Ge}}, \bibinfo {author} {\bibfnamefont {J.-F.}\ \bibnamefont {Wang}}, \bibinfo {author} {\bibfnamefont {Q.}~\bibnamefont {Li}}, \bibinfo {author} {\bibfnamefont {J.-S.}\ \bibnamefont {Xu}}, \bibinfo {author} {\bibfnamefont {Y.-T.}\ \bibnamefont {Wang}}, \bibinfo {author} {\bibfnamefont {J.-S.}\ \bibnamefont {Tang}}, \bibinfo
  {author} {\bibfnamefont {A.}~\bibnamefont {Gali}}, \bibinfo {author} {\bibfnamefont {C.-F.}\ \bibnamefont {Li}},\ and\ \bibinfo {author} {\bibfnamefont {G.-C.}\ \bibnamefont {Guo}},\ }\bibfield  {title} {\bibinfo {title} {Coherent control of an ultrabright single spin in hexagonal boron nitride at room temperature},\ }\href {https://doi.org/10.1038/s41467-023-38672-6} {\bibfield  {journal} {\bibinfo  {journal} {Nature Communications}\ }\textbf {\bibinfo {volume} {14}},\ \bibinfo {pages} {2893} (\bibinfo {year} {2023})}\BibitemShut {NoStop}%
\bibitem [{\citenamefont {Chejanovsky}\ \emph {et~al.}(2021)\citenamefont {Chejanovsky}, \citenamefont {Mukherjee}, \citenamefont {Geng}, \citenamefont {Chen}, \citenamefont {Kim}, \citenamefont {Denisenko}, \citenamefont {Finkler}, \citenamefont {Taniguchi}, \citenamefont {Watanabe}, \citenamefont {Dasari}, \citenamefont {Auburger}, \citenamefont {Gali}, \citenamefont {Smet},\ and\ \citenamefont {Wrachtrup}}]{chejanovskySinglespinResonanceVan2021}%
  \BibitemOpen
  \bibfield  {author} {\bibinfo {author} {\bibfnamefont {N.}~\bibnamefont {Chejanovsky}}, \bibinfo {author} {\bibfnamefont {A.}~\bibnamefont {Mukherjee}}, \bibinfo {author} {\bibfnamefont {J.}~\bibnamefont {Geng}}, \bibinfo {author} {\bibfnamefont {Y.-C.}\ \bibnamefont {Chen}}, \bibinfo {author} {\bibfnamefont {Y.}~\bibnamefont {Kim}}, \bibinfo {author} {\bibfnamefont {A.}~\bibnamefont {Denisenko}}, \bibinfo {author} {\bibfnamefont {A.}~\bibnamefont {Finkler}}, \bibinfo {author} {\bibfnamefont {T.}~\bibnamefont {Taniguchi}}, \bibinfo {author} {\bibfnamefont {K.}~\bibnamefont {Watanabe}}, \bibinfo {author} {\bibfnamefont {D.~B.~R.}\ \bibnamefont {Dasari}}, \bibinfo {author} {\bibfnamefont {P.}~\bibnamefont {Auburger}}, \bibinfo {author} {\bibfnamefont {A.}~\bibnamefont {Gali}}, \bibinfo {author} {\bibfnamefont {J.~H.}\ \bibnamefont {Smet}},\ and\ \bibinfo {author} {\bibfnamefont {J.}~\bibnamefont {Wrachtrup}},\ }\bibfield  {title} {\bibinfo {title} {Single-spin resonance in a van der waals embedded
  paramagnetic defect},\ }\bibfield  {journal} {\bibinfo  {journal} {Nature Materials}\ }\href {https://doi.org/10.1038/s41563-021-00979-4} {10.1038/s41563-021-00979-4} (\bibinfo {year} {2021})\BibitemShut {NoStop}%
\bibitem [{\citenamefont {Patel}\ \emph {et~al.}(2023)\citenamefont {Patel}, \citenamefont {Fishman}, \citenamefont {Huang}, \citenamefont {Gusdorff}, \citenamefont {Fehr}, \citenamefont {Hopper}, \citenamefont {Breitweiser}, \citenamefont {Porat}, \citenamefont {Flatt{\'e}},\ and\ \citenamefont {Bassett}}]{patelRoomTemperatureDynamics2023}%
  \BibitemOpen
  \bibfield  {author} {\bibinfo {author} {\bibfnamefont {R.~N.}\ \bibnamefont {Patel}}, \bibinfo {author} {\bibfnamefont {R.~E.~K.}\ \bibnamefont {Fishman}}, \bibinfo {author} {\bibfnamefont {T.-Y.}\ \bibnamefont {Huang}}, \bibinfo {author} {\bibfnamefont {J.~A.}\ \bibnamefont {Gusdorff}}, \bibinfo {author} {\bibfnamefont {D.~A.}\ \bibnamefont {Fehr}}, \bibinfo {author} {\bibfnamefont {D.~A.}\ \bibnamefont {Hopper}}, \bibinfo {author} {\bibfnamefont {S.~A.}\ \bibnamefont {Breitweiser}}, \bibinfo {author} {\bibfnamefont {B.}~\bibnamefont {Porat}}, \bibinfo {author} {\bibfnamefont {M.~E.}\ \bibnamefont {Flatt{\'e}}},\ and\ \bibinfo {author} {\bibfnamefont {L.~C.}\ \bibnamefont {Bassett}},\ }\href {https://doi.org/10.48550/arXiv.2309.05604} {\bibinfo {title} {Room {{Temperature Dynamics}} of an {{Optically Addressable Single Spin}} in {{Hexagonal Boron Nitride}}}} (\bibinfo {year} {2023}),\ \Eprint {https://arxiv.org/abs/2309.05604} {arXiv:2309.05604} \BibitemShut {NoStop}%
\bibitem [{\citenamefont {Stern}\ \emph {et~al.}(2023)\citenamefont {Stern}, \citenamefont {Gilardoni}, \citenamefont {Gu}, \citenamefont {Barker}, \citenamefont {Powell}, \citenamefont {Deng}, \citenamefont {Follet}, \citenamefont {Li}, \citenamefont {Ramsay}, \citenamefont {Tan}, \citenamefont {Aharonovich},\ and\ \citenamefont {Atat{\"u}re}}]{sternQuantumCoherentSpin2023}%
  \BibitemOpen
  \bibfield  {author} {\bibinfo {author} {\bibfnamefont {H.~L.}\ \bibnamefont {Stern}}, \bibinfo {author} {\bibfnamefont {C.~M.}\ \bibnamefont {Gilardoni}}, \bibinfo {author} {\bibfnamefont {Q.}~\bibnamefont {Gu}}, \bibinfo {author} {\bibfnamefont {S.~E.}\ \bibnamefont {Barker}}, \bibinfo {author} {\bibfnamefont {O.}~\bibnamefont {Powell}}, \bibinfo {author} {\bibfnamefont {X.}~\bibnamefont {Deng}}, \bibinfo {author} {\bibfnamefont {L.}~\bibnamefont {Follet}}, \bibinfo {author} {\bibfnamefont {C.}~\bibnamefont {Li}}, \bibinfo {author} {\bibfnamefont {A.}~\bibnamefont {Ramsay}}, \bibinfo {author} {\bibfnamefont {H.~H.}\ \bibnamefont {Tan}}, \bibinfo {author} {\bibfnamefont {I.}~\bibnamefont {Aharonovich}},\ and\ \bibinfo {author} {\bibfnamefont {M.}~\bibnamefont {Atat{\"u}re}},\ }\href {https://doi.org/10.48550/arXiv.2306.13025} {\bibinfo {title} {A quantum coherent spin in a two-dimensional material at room temperature}} (\bibinfo {year} {2023}),\ \Eprint {https://arxiv.org/abs/2306.13025} {arXiv:2306.13025
  [cond-mat, physics:quant-ph]} \BibitemShut {NoStop}%
\bibitem [{\citenamefont {Stern}\ \emph {et~al.}(2022)\citenamefont {Stern}, \citenamefont {Gu}, \citenamefont {Jarman}, \citenamefont {Eizagirre~Barker}, \citenamefont {Mendelson}, \citenamefont {Chugh}, \citenamefont {Schott}, \citenamefont {Tan}, \citenamefont {Sirringhaus}, \citenamefont {Aharonovich},\ and\ \citenamefont {Atat{\"u}re}}]{sternRoomtemperatureOpticallyDetected2022}%
  \BibitemOpen
  \bibfield  {author} {\bibinfo {author} {\bibfnamefont {H.~L.}\ \bibnamefont {Stern}}, \bibinfo {author} {\bibfnamefont {Q.}~\bibnamefont {Gu}}, \bibinfo {author} {\bibfnamefont {J.}~\bibnamefont {Jarman}}, \bibinfo {author} {\bibfnamefont {S.}~\bibnamefont {Eizagirre~Barker}}, \bibinfo {author} {\bibfnamefont {N.}~\bibnamefont {Mendelson}}, \bibinfo {author} {\bibfnamefont {D.}~\bibnamefont {Chugh}}, \bibinfo {author} {\bibfnamefont {S.}~\bibnamefont {Schott}}, \bibinfo {author} {\bibfnamefont {H.~H.}\ \bibnamefont {Tan}}, \bibinfo {author} {\bibfnamefont {H.}~\bibnamefont {Sirringhaus}}, \bibinfo {author} {\bibfnamefont {I.}~\bibnamefont {Aharonovich}},\ and\ \bibinfo {author} {\bibfnamefont {M.}~\bibnamefont {Atat{\"u}re}},\ }\bibfield  {title} {\bibinfo {title} {Room-temperature optically detected magnetic resonance of single defects in hexagonal boron nitride},\ }\href {https://doi.org/10.1038/s41467-022-28169-z} {\bibfield  {journal} {\bibinfo  {journal} {Nature Communications}\ }\textbf {\bibinfo
  {volume} {13}},\ \bibinfo {pages} {618} (\bibinfo {year} {2022})}\BibitemShut {NoStop}%
\bibitem [{\citenamefont {Mendelson}\ \emph {et~al.}(2021)\citenamefont {Mendelson}, \citenamefont {Chugh}, \citenamefont {Reimers}, \citenamefont {Cheng}, \citenamefont {Gottscholl}, \citenamefont {Long}, \citenamefont {Mellor}, \citenamefont {Zettl}, \citenamefont {Dyakonov}, \citenamefont {Beton}, \citenamefont {Novikov}, \citenamefont {Jagadish}, \citenamefont {Tan}, \citenamefont {Ford}, \citenamefont {Toth}, \citenamefont {Bradac},\ and\ \citenamefont {Aharonovich}}]{mendelsonIdentifyingCarbonSource2021}%
  \BibitemOpen
  \bibfield  {author} {\bibinfo {author} {\bibfnamefont {N.}~\bibnamefont {Mendelson}}, \bibinfo {author} {\bibfnamefont {D.}~\bibnamefont {Chugh}}, \bibinfo {author} {\bibfnamefont {J.~R.}\ \bibnamefont {Reimers}}, \bibinfo {author} {\bibfnamefont {T.~S.}\ \bibnamefont {Cheng}}, \bibinfo {author} {\bibfnamefont {A.}~\bibnamefont {Gottscholl}}, \bibinfo {author} {\bibfnamefont {H.}~\bibnamefont {Long}}, \bibinfo {author} {\bibfnamefont {C.~J.}\ \bibnamefont {Mellor}}, \bibinfo {author} {\bibfnamefont {A.}~\bibnamefont {Zettl}}, \bibinfo {author} {\bibfnamefont {V.}~\bibnamefont {Dyakonov}}, \bibinfo {author} {\bibfnamefont {P.~H.}\ \bibnamefont {Beton}}, \bibinfo {author} {\bibfnamefont {S.~V.}\ \bibnamefont {Novikov}}, \bibinfo {author} {\bibfnamefont {C.}~\bibnamefont {Jagadish}}, \bibinfo {author} {\bibfnamefont {H.~H.}\ \bibnamefont {Tan}}, \bibinfo {author} {\bibfnamefont {M.~J.}\ \bibnamefont {Ford}}, \bibinfo {author} {\bibfnamefont {M.}~\bibnamefont {Toth}}, \bibinfo {author} {\bibfnamefont
  {C.}~\bibnamefont {Bradac}},\ and\ \bibinfo {author} {\bibfnamefont {I.}~\bibnamefont {Aharonovich}},\ }\bibfield  {title} {\bibinfo {title} {Identifying carbon as the source of visible single-photon emission from hexagonal boron nitride},\ }\href {https://doi.org/10.1038/s41563-020-00850-y} {\bibfield  {journal} {\bibinfo  {journal} {Nature Materials}\ }\textbf {\bibinfo {volume} {20}},\ \bibinfo {pages} {321} (\bibinfo {year} {2021})}\BibitemShut {NoStop}%
\bibitem [{\citenamefont {Scholten}\ \emph {et~al.}(2023)\citenamefont {Scholten}, \citenamefont {Singh}, \citenamefont {Healey}, \citenamefont {Robertson}, \citenamefont {Haim}, \citenamefont {Tan}, \citenamefont {Broadway}, \citenamefont {Wang}, \citenamefont {Abe}, \citenamefont {Ohshima}, \citenamefont {Kianinia}, \citenamefont {Reineck}, \citenamefont {Aharonovich},\ and\ \citenamefont {Tetienne}}]{scholtenMultispeciesOpticallyAddressable2023}%
  \BibitemOpen
  \bibfield  {author} {\bibinfo {author} {\bibfnamefont {S.~C.}\ \bibnamefont {Scholten}}, \bibinfo {author} {\bibfnamefont {P.}~\bibnamefont {Singh}}, \bibinfo {author} {\bibfnamefont {A.~J.}\ \bibnamefont {Healey}}, \bibinfo {author} {\bibfnamefont {I.~O.}\ \bibnamefont {Robertson}}, \bibinfo {author} {\bibfnamefont {G.}~\bibnamefont {Haim}}, \bibinfo {author} {\bibfnamefont {C.}~\bibnamefont {Tan}}, \bibinfo {author} {\bibfnamefont {D.~A.}\ \bibnamefont {Broadway}}, \bibinfo {author} {\bibfnamefont {L.}~\bibnamefont {Wang}}, \bibinfo {author} {\bibfnamefont {H.}~\bibnamefont {Abe}}, \bibinfo {author} {\bibfnamefont {T.}~\bibnamefont {Ohshima}}, \bibinfo {author} {\bibfnamefont {M.}~\bibnamefont {Kianinia}}, \bibinfo {author} {\bibfnamefont {P.}~\bibnamefont {Reineck}}, \bibinfo {author} {\bibfnamefont {I.}~\bibnamefont {Aharonovich}},\ and\ \bibinfo {author} {\bibfnamefont {J.-P.}\ \bibnamefont {Tetienne}},\ }\href {https://doi.org/10.48550/arXiv.2306.16600} {\bibinfo {title} {Multi-species optically
  addressable spin defects in a van der {{Waals}} material}} (\bibinfo {year} {2023}),\ \Eprint {https://arxiv.org/abs/2306.16600} {arXiv:2306.16600 [cond-mat]} \BibitemShut {NoStop}%
\bibitem [{\citenamefont {Singh}\ \emph {et~al.}(2025)\citenamefont {Singh}, \citenamefont {Robertson}, \citenamefont {Scholten}, \citenamefont {Healey}, \citenamefont {Abe}, \citenamefont {Ohshima}, \citenamefont {Tan}, \citenamefont {Kianinia}, \citenamefont {Aharonovich}, \citenamefont {Broadway}, \citenamefont {Reineck},\ and\ \citenamefont {Tetienne}}]{singhVioletNearInfraredOptical2025}%
  \BibitemOpen
  \bibfield  {author} {\bibinfo {author} {\bibfnamefont {P.}~\bibnamefont {Singh}}, \bibinfo {author} {\bibfnamefont {I.~O.}\ \bibnamefont {Robertson}}, \bibinfo {author} {\bibfnamefont {S.~C.}\ \bibnamefont {Scholten}}, \bibinfo {author} {\bibfnamefont {A.~J.}\ \bibnamefont {Healey}}, \bibinfo {author} {\bibfnamefont {H.}~\bibnamefont {Abe}}, \bibinfo {author} {\bibfnamefont {T.}~\bibnamefont {Ohshima}}, \bibinfo {author} {\bibfnamefont {H.~H.}\ \bibnamefont {Tan}}, \bibinfo {author} {\bibfnamefont {M.}~\bibnamefont {Kianinia}}, \bibinfo {author} {\bibfnamefont {I.}~\bibnamefont {Aharonovich}}, \bibinfo {author} {\bibfnamefont {D.~A.}\ \bibnamefont {Broadway}}, \bibinfo {author} {\bibfnamefont {P.}~\bibnamefont {Reineck}},\ and\ \bibinfo {author} {\bibfnamefont {J.-P.}\ \bibnamefont {Tetienne}},\ }\bibfield  {title} {\bibinfo {title} {Violet to {{Near-Infrared Optical Addressing}} of {{Spin Pairs}} in {{Hexagonal Boron Nitride}}},\ }\href {https://doi.org/10.1002/adma.202414846} {\bibfield  {journal}
  {\bibinfo  {journal} {Advanced Materials}\ }\textbf {\bibinfo {volume} {n/a}},\ \bibinfo {pages} {2414846} (\bibinfo {year} {2025})}\BibitemShut {NoStop}%
\bibitem [{\citenamefont {Gao}\ \emph {et~al.}(2024)\citenamefont {Gao}, \citenamefont {Vaidya}, \citenamefont {Li}, \citenamefont {Dikshit}, \citenamefont {Zhang}, \citenamefont {Ju}, \citenamefont {Shen}, \citenamefont {Jin}, \citenamefont {Ping},\ and\ \citenamefont {Li}}]{gaoSingleNuclearSpin2024}%
  \BibitemOpen
  \bibfield  {author} {\bibinfo {author} {\bibfnamefont {X.}~\bibnamefont {Gao}}, \bibinfo {author} {\bibfnamefont {S.}~\bibnamefont {Vaidya}}, \bibinfo {author} {\bibfnamefont {K.}~\bibnamefont {Li}}, \bibinfo {author} {\bibfnamefont {S.}~\bibnamefont {Dikshit}}, \bibinfo {author} {\bibfnamefont {S.}~\bibnamefont {Zhang}}, \bibinfo {author} {\bibfnamefont {P.}~\bibnamefont {Ju}}, \bibinfo {author} {\bibfnamefont {K.}~\bibnamefont {Shen}}, \bibinfo {author} {\bibfnamefont {Y.}~\bibnamefont {Jin}}, \bibinfo {author} {\bibfnamefont {Y.}~\bibnamefont {Ping}},\ and\ \bibinfo {author} {\bibfnamefont {T.}~\bibnamefont {Li}},\ }\href {https://doi.org/10.48550/arXiv.2409.01601} {\bibinfo {title} {Single nuclear spin detection and control in a van der {{Waals}} material}} (\bibinfo {year} {2024}),\ \Eprint {https://arxiv.org/abs/2409.01601} {arXiv:2409.01601 [cond-mat, physics:quant-ph]} \BibitemShut {NoStop}%
\bibitem [{\citenamefont {Whitefield}\ \emph {et~al.}(2025)\citenamefont {Whitefield}, \citenamefont {Zeng}, \citenamefont {{Liddle-Wesolowski}}, \citenamefont {Robertson}, \citenamefont {Iv{\'a}dy}, \citenamefont {Watanabe}, \citenamefont {Taniguchi}, \citenamefont {Toth}, \citenamefont {Tetienne}, \citenamefont {Aharonovich},\ and\ \citenamefont {Kianinia}}]{whitefieldGenerationNarrowbandQuantum2025}%
  \BibitemOpen
  \bibfield  {author} {\bibinfo {author} {\bibfnamefont {B.}~\bibnamefont {Whitefield}}, \bibinfo {author} {\bibfnamefont {H.~Z.~J.}\ \bibnamefont {Zeng}}, \bibinfo {author} {\bibfnamefont {J.}~\bibnamefont {{Liddle-Wesolowski}}}, \bibinfo {author} {\bibfnamefont {I.~O.}\ \bibnamefont {Robertson}}, \bibinfo {author} {\bibfnamefont {V.}~\bibnamefont {Iv{\'a}dy}}, \bibinfo {author} {\bibfnamefont {K.}~\bibnamefont {Watanabe}}, \bibinfo {author} {\bibfnamefont {T.}~\bibnamefont {Taniguchi}}, \bibinfo {author} {\bibfnamefont {M.}~\bibnamefont {Toth}}, \bibinfo {author} {\bibfnamefont {J.-P.}\ \bibnamefont {Tetienne}}, \bibinfo {author} {\bibfnamefont {I.}~\bibnamefont {Aharonovich}},\ and\ \bibinfo {author} {\bibfnamefont {M.}~\bibnamefont {Kianinia}},\ }\href {https://doi.org/10.48550/arXiv.2501.15341} {\bibinfo {title} {Generation of narrowband quantum emitters in {{hBN}} with optically addressable spins}} (\bibinfo {year} {2025}),\ \Eprint {https://arxiv.org/abs/2501.15341} {arXiv:2501.15341 [quant-ph]}
  \BibitemShut {NoStop}%
\bibitem [{\citenamefont {Gilardoni}\ \emph {et~al.}(2024)\citenamefont {Gilardoni}, \citenamefont {Barker}, \citenamefont {Curtin}, \citenamefont {Fraser}, \citenamefont {Powell}, \citenamefont {Lewis}, \citenamefont {Deng}, \citenamefont {Ramsay}, \citenamefont {Li}, \citenamefont {Aharonovich}, \citenamefont {Tan}, \citenamefont {Atat{\"u}re},\ and\ \citenamefont {Stern}}]{gilardoniSingleSpinHexagonal2024}%
  \BibitemOpen
  \bibfield  {author} {\bibinfo {author} {\bibfnamefont {C.~M.}\ \bibnamefont {Gilardoni}}, \bibinfo {author} {\bibfnamefont {S.~E.}\ \bibnamefont {Barker}}, \bibinfo {author} {\bibfnamefont {C.~L.}\ \bibnamefont {Curtin}}, \bibinfo {author} {\bibfnamefont {S.~A.}\ \bibnamefont {Fraser}}, \bibinfo {author} {\bibfnamefont {O.~F.~J.}\ \bibnamefont {Powell}}, \bibinfo {author} {\bibfnamefont {D.~K.}\ \bibnamefont {Lewis}}, \bibinfo {author} {\bibfnamefont {X.}~\bibnamefont {Deng}}, \bibinfo {author} {\bibfnamefont {A.~J.}\ \bibnamefont {Ramsay}}, \bibinfo {author} {\bibfnamefont {C.}~\bibnamefont {Li}}, \bibinfo {author} {\bibfnamefont {I.}~\bibnamefont {Aharonovich}}, \bibinfo {author} {\bibfnamefont {H.~H.}\ \bibnamefont {Tan}}, \bibinfo {author} {\bibfnamefont {M.}~\bibnamefont {Atat{\"u}re}},\ and\ \bibinfo {author} {\bibfnamefont {H.~L.}\ \bibnamefont {Stern}},\ }\href {https://doi.org/10.48550/arXiv.2408.10348} {\bibinfo {title} {A single spin in hexagonal boron nitride for vectorial quantum magnetometry}}
  (\bibinfo {year} {2024}),\ \Eprint {https://arxiv.org/abs/2408.10348} {arXiv:2408.10348 [cond-mat, physics:quant-ph]} \BibitemShut {NoStop}%
\bibitem [{\citenamefont {Robertson}\ \emph {et~al.}(2024)\citenamefont {Robertson}, \citenamefont {Whitefield}, \citenamefont {Scholten}, \citenamefont {Singh}, \citenamefont {Healey}, \citenamefont {Reineck}, \citenamefont {Kianinia}, \citenamefont {Broadway}, \citenamefont {Aharonovich},\ and\ \citenamefont {Tetienne}}]{robertsonUniversalMechanismOptically2024}%
  \BibitemOpen
  \bibfield  {author} {\bibinfo {author} {\bibfnamefont {I.~O.}\ \bibnamefont {Robertson}}, \bibinfo {author} {\bibfnamefont {B.}~\bibnamefont {Whitefield}}, \bibinfo {author} {\bibfnamefont {S.~C.}\ \bibnamefont {Scholten}}, \bibinfo {author} {\bibfnamefont {P.}~\bibnamefont {Singh}}, \bibinfo {author} {\bibfnamefont {A.~J.}\ \bibnamefont {Healey}}, \bibinfo {author} {\bibfnamefont {P.}~\bibnamefont {Reineck}}, \bibinfo {author} {\bibfnamefont {M.}~\bibnamefont {Kianinia}}, \bibinfo {author} {\bibfnamefont {D.~A.}\ \bibnamefont {Broadway}}, \bibinfo {author} {\bibfnamefont {I.}~\bibnamefont {Aharonovich}},\ and\ \bibinfo {author} {\bibfnamefont {J.-P.}\ \bibnamefont {Tetienne}},\ }\href {https://doi.org/10.48550/arXiv.2407.13148} {\bibinfo {title} {A universal mechanism for optically addressable solid-state spin pairs}} (\bibinfo {year} {2024}),\ \Eprint {https://arxiv.org/abs/2407.13148} {arXiv:2407.13148 [cond-mat]} \BibitemShut {NoStop}%
\bibitem [{\citenamefont {Yang}\ \emph {et~al.}(2023)\citenamefont {Yang}, \citenamefont {Zhu}, \citenamefont {Li}, \citenamefont {Zeng}, \citenamefont {Guo}, \citenamefont {Yu}, \citenamefont {Meng}, \citenamefont {Wang}, \citenamefont {Xie}, \citenamefont {Zhou}, \citenamefont {Li}, \citenamefont {Xu}, \citenamefont {Gao}, \citenamefont {Liu}, \citenamefont {Wang}, \citenamefont {Tang}, \citenamefont {Li},\ and\ \citenamefont {Guo}}]{yangLaserDirectWriting2023}%
  \BibitemOpen
  \bibfield  {author} {\bibinfo {author} {\bibfnamefont {Y.-Z.}\ \bibnamefont {Yang}}, \bibinfo {author} {\bibfnamefont {T.-X.}\ \bibnamefont {Zhu}}, \bibinfo {author} {\bibfnamefont {Z.-P.}\ \bibnamefont {Li}}, \bibinfo {author} {\bibfnamefont {X.-D.}\ \bibnamefont {Zeng}}, \bibinfo {author} {\bibfnamefont {N.-J.}\ \bibnamefont {Guo}}, \bibinfo {author} {\bibfnamefont {S.}~\bibnamefont {Yu}}, \bibinfo {author} {\bibfnamefont {Y.}~\bibnamefont {Meng}}, \bibinfo {author} {\bibfnamefont {Z.-A.}\ \bibnamefont {Wang}}, \bibinfo {author} {\bibfnamefont {L.-K.}\ \bibnamefont {Xie}}, \bibinfo {author} {\bibfnamefont {Z.-Q.}\ \bibnamefont {Zhou}}, \bibinfo {author} {\bibfnamefont {Q.}~\bibnamefont {Li}}, \bibinfo {author} {\bibfnamefont {J.-S.}\ \bibnamefont {Xu}}, \bibinfo {author} {\bibfnamefont {X.-Y.}\ \bibnamefont {Gao}}, \bibinfo {author} {\bibfnamefont {W.}~\bibnamefont {Liu}}, \bibinfo {author} {\bibfnamefont {Y.-T.}\ \bibnamefont {Wang}}, \bibinfo {author} {\bibfnamefont {J.-S.}\ \bibnamefont {Tang}},
  \bibinfo {author} {\bibfnamefont {C.-F.}\ \bibnamefont {Li}},\ and\ \bibinfo {author} {\bibfnamefont {G.-C.}\ \bibnamefont {Guo}},\ }\bibfield  {title} {\bibinfo {title} {Laser {{Direct Writing}} of {{Visible Spin Defects}} in {{Hexagonal Boron Nitride}} for {{Applications}} in {{Spin-Based Technologies}}},\ }\href {https://doi.org/10.1021/acsanm.3c01047} {\bibfield  {journal} {\bibinfo  {journal} {ACS Applied Nano Materials}\ }\textbf {\bibinfo {volume} {6}},\ \bibinfo {pages} {6407} (\bibinfo {year} {2023})}\BibitemShut {NoStop}%
\bibitem [{\citenamefont {D{\k a}browska}\ \emph {et~al.}(2024)\citenamefont {D{\k a}browska}, \citenamefont {Binder}, \citenamefont {Prozheev}, \citenamefont {Tuomisto}, \citenamefont {Iwa{\'n}ski}, \citenamefont {Tokarczyk}, \citenamefont {Korona}, \citenamefont {Kowalski}, \citenamefont {St{\k e}pniewski},\ and\ \citenamefont {Wysmo{\l}ek}}]{dabrowskaDefectsLayeredBoron2024}%
  \BibitemOpen
  \bibfield  {author} {\bibinfo {author} {\bibfnamefont {A.~K.}\ \bibnamefont {D{\k a}browska}}, \bibinfo {author} {\bibfnamefont {J.}~\bibnamefont {Binder}}, \bibinfo {author} {\bibfnamefont {I.}~\bibnamefont {Prozheev}}, \bibinfo {author} {\bibfnamefont {F.}~\bibnamefont {Tuomisto}}, \bibinfo {author} {\bibfnamefont {J.}~\bibnamefont {Iwa{\'n}ski}}, \bibinfo {author} {\bibfnamefont {M.}~\bibnamefont {Tokarczyk}}, \bibinfo {author} {\bibfnamefont {K.~P.}\ \bibnamefont {Korona}}, \bibinfo {author} {\bibfnamefont {G.}~\bibnamefont {Kowalski}}, \bibinfo {author} {\bibfnamefont {R.}~\bibnamefont {St{\k e}pniewski}},\ and\ \bibinfo {author} {\bibfnamefont {A.}~\bibnamefont {Wysmo{\l}ek}},\ }\bibfield  {title} {\bibinfo {title} {Defects in layered boron nitride grown by {{Metal Organic Vapor Phase Epitaxy}}: Luminescence and positron annihilation studies},\ }\href {https://doi.org/10.1016/j.jlumin.2024.120486} {\bibfield  {journal} {\bibinfo  {journal} {Journal of Luminescence}\ }\textbf {\bibinfo {volume}
  {269}},\ \bibinfo {pages} {120486} (\bibinfo {year} {2024})}\BibitemShut {NoStop}%
\bibitem [{\citenamefont {Iwa{\'n}ski}\ \emph {et~al.}(2024)\citenamefont {Iwa{\'n}ski}, \citenamefont {Kierdaszuk}, \citenamefont {Ciesielski}, \citenamefont {Binder}, \citenamefont {Drabi{\'n}ska},\ and\ \citenamefont {Wysmo{\l}ek}}]{iwanskiManipulatingCarbonRelated2024}%
  \BibitemOpen
  \bibfield  {author} {\bibinfo {author} {\bibfnamefont {J.}~\bibnamefont {Iwa{\'n}ski}}, \bibinfo {author} {\bibfnamefont {J.}~\bibnamefont {Kierdaszuk}}, \bibinfo {author} {\bibfnamefont {A.}~\bibnamefont {Ciesielski}}, \bibinfo {author} {\bibfnamefont {J.}~\bibnamefont {Binder}}, \bibinfo {author} {\bibfnamefont {A.}~\bibnamefont {Drabi{\'n}ska}},\ and\ \bibinfo {author} {\bibfnamefont {A.}~\bibnamefont {Wysmo{\l}ek}},\ }\bibfield  {title} {\bibinfo {title} {Manipulating carbon related spin defects in boron nitride by changing the {{MOCVD}} growth temperature},\ }\href {https://doi.org/10.1016/j.diamond.2024.111291} {\bibfield  {journal} {\bibinfo  {journal} {Diamond and Related Materials}\ }\textbf {\bibinfo {volume} {147}},\ \bibinfo {pages} {111291} (\bibinfo {year} {2024})}\BibitemShut {NoStop}%
\bibitem [{\citenamefont {Hua}\ \emph {et~al.}(2024)\citenamefont {Hua}, \citenamefont {Chen}, \citenamefont {Hou}, \citenamefont {Kolluru}, \citenamefont {Chan}, \citenamefont {Liu}, \citenamefont {Gage}, \citenamefont {Zuo}, \citenamefont {Diroll},\ and\ \citenamefont {Wen}}]{huaDeterministicCreationIdentical2024}%
  \BibitemOpen
  \bibfield  {author} {\bibinfo {author} {\bibfnamefont {M.}~\bibnamefont {Hua}}, \bibinfo {author} {\bibfnamefont {W.-Y.}\ \bibnamefont {Chen}}, \bibinfo {author} {\bibfnamefont {H.}~\bibnamefont {Hou}}, \bibinfo {author} {\bibfnamefont {V.~S.~C.}\ \bibnamefont {Kolluru}}, \bibinfo {author} {\bibfnamefont {M.~K.~Y.}\ \bibnamefont {Chan}}, \bibinfo {author} {\bibfnamefont {H.}~\bibnamefont {Liu}}, \bibinfo {author} {\bibfnamefont {T.~E.}\ \bibnamefont {Gage}}, \bibinfo {author} {\bibfnamefont {J.-M.}\ \bibnamefont {Zuo}}, \bibinfo {author} {\bibfnamefont {B.~T.}\ \bibnamefont {Diroll}},\ and\ \bibinfo {author} {\bibfnamefont {J.}~\bibnamefont {Wen}},\ }\href {https://doi.org/10.48550/arXiv.2410.13169} {\bibinfo {title} {Deterministic {{Creation}} of {{Identical Monochromatic Quantum Emitters}} in {{Hexagonal Boron Nitride}}}} (\bibinfo {year} {2024}),\ \Eprint {https://arxiv.org/abs/2410.13169} {arXiv:2410.13169} \BibitemShut {NoStop}%
\bibitem [{\citenamefont {Zhigulin}\ \emph {et~al.}(2025)\citenamefont {Zhigulin}, \citenamefont {Park}, \citenamefont {Yamamura}, \citenamefont {Watanabe}, \citenamefont {Taniguchi}, \citenamefont {Toth}, \citenamefont {Kim},\ and\ \citenamefont {Aharonovich}}]{zhigulinElectricalGenerationColour2025}%
  \BibitemOpen
  \bibfield  {author} {\bibinfo {author} {\bibfnamefont {I.}~\bibnamefont {Zhigulin}}, \bibinfo {author} {\bibfnamefont {G.}~\bibnamefont {Park}}, \bibinfo {author} {\bibfnamefont {K.}~\bibnamefont {Yamamura}}, \bibinfo {author} {\bibfnamefont {K.}~\bibnamefont {Watanabe}}, \bibinfo {author} {\bibfnamefont {T.}~\bibnamefont {Taniguchi}}, \bibinfo {author} {\bibfnamefont {M.}~\bibnamefont {Toth}}, \bibinfo {author} {\bibfnamefont {J.}~\bibnamefont {Kim}},\ and\ \bibinfo {author} {\bibfnamefont {I.}~\bibnamefont {Aharonovich}},\ }\href {https://doi.org/10.48550/arXiv.2501.07846} {\bibinfo {title} {Electrical {{Generation}} of {{Colour Centres}} in {{Hexagonal Boron Nitride}}}} (\bibinfo {year} {2025}),\ \Eprint {https://arxiv.org/abs/2501.07846} {arXiv:2501.07846 [physics]} \BibitemShut {NoStop}%
\bibitem [{\citenamefont {Stern}\ \emph {et~al.}(2019)\citenamefont {Stern}, \citenamefont {Wang}, \citenamefont {Fan}, \citenamefont {Mizuta}, \citenamefont {Stewart}, \citenamefont {Needham}, \citenamefont {Roberts}, \citenamefont {Wai}, \citenamefont {Ginsberg}, \citenamefont {Klenerman}, \citenamefont {Hofmann},\ and\ \citenamefont {Lee}}]{sternSpectrallyResolvedPhotodynamics2019}%
  \BibitemOpen
  \bibfield  {author} {\bibinfo {author} {\bibfnamefont {H.~L.}\ \bibnamefont {Stern}}, \bibinfo {author} {\bibfnamefont {R.}~\bibnamefont {Wang}}, \bibinfo {author} {\bibfnamefont {Y.}~\bibnamefont {Fan}}, \bibinfo {author} {\bibfnamefont {R.}~\bibnamefont {Mizuta}}, \bibinfo {author} {\bibfnamefont {J.~C.}\ \bibnamefont {Stewart}}, \bibinfo {author} {\bibfnamefont {L.-M.}\ \bibnamefont {Needham}}, \bibinfo {author} {\bibfnamefont {T.~D.}\ \bibnamefont {Roberts}}, \bibinfo {author} {\bibfnamefont {R.}~\bibnamefont {Wai}}, \bibinfo {author} {\bibfnamefont {N.~S.}\ \bibnamefont {Ginsberg}}, \bibinfo {author} {\bibfnamefont {D.}~\bibnamefont {Klenerman}}, \bibinfo {author} {\bibfnamefont {S.}~\bibnamefont {Hofmann}},\ and\ \bibinfo {author} {\bibfnamefont {S.~F.}\ \bibnamefont {Lee}},\ }\bibfield  {title} {\bibinfo {title} {Spectrally {{Resolved Photodynamics}} of {{Individual Emitters}} in {{Large-Area Monolayers}} of {{Hexagonal Boron Nitride}}},\ }\href {https://doi.org/10.1021/acsnano.9b00274} {\bibfield
  {journal} {\bibinfo  {journal} {ACS Nano}\ }\textbf {\bibinfo {volume} {13}},\ \bibinfo {pages} {4538} (\bibinfo {year} {2019})}\BibitemShut {NoStop}%
\bibitem [{\citenamefont {Chugh}\ \emph {et~al.}(2018)\citenamefont {Chugh}, \citenamefont {{Wong-Leung}}, \citenamefont {Li}, \citenamefont {Lysevych}, \citenamefont {Tan},\ and\ \citenamefont {Jagadish}}]{chughFlowModulationEpitaxy2018}%
  \BibitemOpen
  \bibfield  {author} {\bibinfo {author} {\bibfnamefont {D.}~\bibnamefont {Chugh}}, \bibinfo {author} {\bibfnamefont {J.}~\bibnamefont {{Wong-Leung}}}, \bibinfo {author} {\bibfnamefont {L.}~\bibnamefont {Li}}, \bibinfo {author} {\bibfnamefont {M.}~\bibnamefont {Lysevych}}, \bibinfo {author} {\bibfnamefont {H.~H.}\ \bibnamefont {Tan}},\ and\ \bibinfo {author} {\bibfnamefont {C.}~\bibnamefont {Jagadish}},\ }\bibfield  {title} {\bibinfo {title} {Flow modulation epitaxy of hexagonal boron nitride},\ }\href {https://doi.org/10.1088/2053-1583/aad5aa} {\bibfield  {journal} {\bibinfo  {journal} {2D Materials}\ }\textbf {\bibinfo {volume} {5}},\ \bibinfo {pages} {045018} (\bibinfo {year} {2018})}\BibitemShut {NoStop}%
\bibitem [{\citenamefont {D{\k a}browska}\ \emph {et~al.}(2020)\citenamefont {D{\k a}browska}, \citenamefont {Tokarczyk}, \citenamefont {Kowalski}, \citenamefont {Binder}, \citenamefont {Bo{\.z}ek}, \citenamefont {Borysiuk}, \citenamefont {St{\k e}pniewski},\ and\ \citenamefont {Wysmo{\l}ek}}]{dabrowskaTwoStageEpitaxial2020}%
  \BibitemOpen
  \bibfield  {author} {\bibinfo {author} {\bibfnamefont {A.~K.}\ \bibnamefont {D{\k a}browska}}, \bibinfo {author} {\bibfnamefont {M.}~\bibnamefont {Tokarczyk}}, \bibinfo {author} {\bibfnamefont {G.}~\bibnamefont {Kowalski}}, \bibinfo {author} {\bibfnamefont {J.}~\bibnamefont {Binder}}, \bibinfo {author} {\bibfnamefont {R.}~\bibnamefont {Bo{\.z}ek}}, \bibinfo {author} {\bibfnamefont {J.}~\bibnamefont {Borysiuk}}, \bibinfo {author} {\bibfnamefont {R.}~\bibnamefont {St{\k e}pniewski}},\ and\ \bibinfo {author} {\bibfnamefont {A.}~\bibnamefont {Wysmo{\l}ek}},\ }\bibfield  {title} {\bibinfo {title} {Two stage epitaxial growth of wafer-size multilayer h-{{BN}} by metal-organic vapor phase epitaxy -- a homoepitaxial approach},\ }\href {https://doi.org/10.1088/2053-1583/abbd1f} {\bibfield  {journal} {\bibinfo  {journal} {2D Materials}\ }\textbf {\bibinfo {volume} {8}},\ \bibinfo {pages} {015017} (\bibinfo {year} {2020})}\BibitemShut {NoStop}%
\bibitem [{\citenamefont {Paku{\l}a}\ \emph {et~al.}(2019)\citenamefont {Paku{\l}a}, \citenamefont {D{\k a}browska}, \citenamefont {Tokarczyk}, \citenamefont {Bo{\.z}ek}, \citenamefont {Binder}, \citenamefont {Kowalski}, \citenamefont {Wysmo{\l}ek},\ and\ \citenamefont {St{\k e}pniewski}}]{pakulaFundamentalMechanismsHBN2019}%
  \BibitemOpen
  \bibfield  {author} {\bibinfo {author} {\bibfnamefont {K.}~\bibnamefont {Paku{\l}a}}, \bibinfo {author} {\bibfnamefont {A.}~\bibnamefont {D{\k a}browska}}, \bibinfo {author} {\bibfnamefont {M.}~\bibnamefont {Tokarczyk}}, \bibinfo {author} {\bibfnamefont {R.}~\bibnamefont {Bo{\.z}ek}}, \bibinfo {author} {\bibfnamefont {J.}~\bibnamefont {Binder}}, \bibinfo {author} {\bibfnamefont {G.}~\bibnamefont {Kowalski}}, \bibinfo {author} {\bibfnamefont {A.}~\bibnamefont {Wysmo{\l}ek}},\ and\ \bibinfo {author} {\bibfnamefont {R.}~\bibnamefont {St{\k e}pniewski}},\ }\href {https://doi.org/10.48550/arXiv.1906.05319} {\bibinfo {title} {Fundamental mechanisms of {{hBN}} growth by {{MOVPE}}}} (\bibinfo {year} {2019}),\ \Eprint {https://arxiv.org/abs/1906.05319} {arXiv:1906.05319 [cond-mat]} \BibitemShut {NoStop}%
\bibitem [{\citenamefont {Iwa{\'n}ski}\ \emph {et~al.}(2022)\citenamefont {Iwa{\'n}ski}, \citenamefont {Tatarczak}, \citenamefont {Tokarczyk}, \citenamefont {D{\c a}browska}, \citenamefont {Paw{\l}owski}, \citenamefont {Binder}, \citenamefont {Kowalski}, \citenamefont {St{\c e}pniewski},\ and\ \citenamefont {Wysmo{\l}ek}}]{iwanskiTemperatureInducedGiant2022}%
  \BibitemOpen
  \bibfield  {author} {\bibinfo {author} {\bibfnamefont {J.}~\bibnamefont {Iwa{\'n}ski}}, \bibinfo {author} {\bibfnamefont {P.}~\bibnamefont {Tatarczak}}, \bibinfo {author} {\bibfnamefont {M.}~\bibnamefont {Tokarczyk}}, \bibinfo {author} {\bibfnamefont {A.~K.}\ \bibnamefont {D{\c a}browska}}, \bibinfo {author} {\bibfnamefont {J.}~\bibnamefont {Paw{\l}owski}}, \bibinfo {author} {\bibfnamefont {J.}~\bibnamefont {Binder}}, \bibinfo {author} {\bibfnamefont {G.}~\bibnamefont {Kowalski}}, \bibinfo {author} {\bibfnamefont {R.}~\bibnamefont {St{\c e}pniewski}},\ and\ \bibinfo {author} {\bibfnamefont {A.}~\bibnamefont {Wysmo{\l}ek}},\ }\bibfield  {title} {\bibinfo {title} {Temperature induced giant shift of phonon energy in epitaxial boron nitride layers},\ }\href {https://doi.org/10.1088/1361-6528/ac9629} {\bibfield  {journal} {\bibinfo  {journal} {Nanotechnology}\ }\textbf {\bibinfo {volume} {34}},\ \bibinfo {pages} {015202} (\bibinfo {year} {2022})}\BibitemShut {NoStop}%
\bibitem [{\citenamefont {Tatarczak}\ \emph {et~al.}(2024)\citenamefont {Tatarczak}, \citenamefont {Iwa{\'n}ski}, \citenamefont {D{\k a}browska}, \citenamefont {Tokarczyk}, \citenamefont {Binder}, \citenamefont {St{\k e}pniewski},\ and\ \citenamefont {Wysmo{\l}ek}}]{tatarczakStrainModulationEpitaxial2024}%
  \BibitemOpen
  \bibfield  {author} {\bibinfo {author} {\bibfnamefont {P.}~\bibnamefont {Tatarczak}}, \bibinfo {author} {\bibfnamefont {J.}~\bibnamefont {Iwa{\'n}ski}}, \bibinfo {author} {\bibfnamefont {A.~K.}\ \bibnamefont {D{\k a}browska}}, \bibinfo {author} {\bibfnamefont {M.}~\bibnamefont {Tokarczyk}}, \bibinfo {author} {\bibfnamefont {J.}~\bibnamefont {Binder}}, \bibinfo {author} {\bibfnamefont {R.}~\bibnamefont {St{\k e}pniewski}},\ and\ \bibinfo {author} {\bibfnamefont {A.}~\bibnamefont {Wysmo{\l}ek}},\ }\bibfield  {title} {\bibinfo {title} {Strain modulation of epitaxial h-{{BN}} on sapphire: The role of wrinkle formation for large-area two-dimensional materials},\ }\href {https://doi.org/10.1088/1361-6528/ad18e6} {\bibfield  {journal} {\bibinfo  {journal} {Nanotechnology}\ }\textbf {\bibinfo {volume} {35}},\ \bibinfo {pages} {175703} (\bibinfo {year} {2024})}\BibitemShut {NoStop}%
\bibitem [{\citenamefont {Ciesielski}\ \emph {et~al.}(2023)\citenamefont {Ciesielski}, \citenamefont {Iwa{\'n}ski}, \citenamefont {Wr{\'o}bel}, \citenamefont {Bo{\.z}ek}, \citenamefont {Kret}, \citenamefont {Turczy{\'n}ski}, \citenamefont {Binder}, \citenamefont {Korona}, \citenamefont {St{\k e}pniewski},\ and\ \citenamefont {Wysmo{\l}ek}}]{ciesielskiAllBNDistributedBragg2023}%
  \BibitemOpen
  \bibfield  {author} {\bibinfo {author} {\bibfnamefont {A.}~\bibnamefont {Ciesielski}}, \bibinfo {author} {\bibfnamefont {J.}~\bibnamefont {Iwa{\'n}ski}}, \bibinfo {author} {\bibfnamefont {P.}~\bibnamefont {Wr{\'o}bel}}, \bibinfo {author} {\bibfnamefont {R.}~\bibnamefont {Bo{\.z}ek}}, \bibinfo {author} {\bibfnamefont {S.}~\bibnamefont {Kret}}, \bibinfo {author} {\bibfnamefont {J.}~\bibnamefont {Turczy{\'n}ski}}, \bibinfo {author} {\bibfnamefont {J.}~\bibnamefont {Binder}}, \bibinfo {author} {\bibfnamefont {K.~P.}\ \bibnamefont {Korona}}, \bibinfo {author} {\bibfnamefont {R.}~\bibnamefont {St{\k e}pniewski}},\ and\ \bibinfo {author} {\bibfnamefont {A.}~\bibnamefont {Wysmo{\l}ek}},\ }\bibfield  {title} {\bibinfo {title} {All-{{BN}} distributed {{Bragg}} reflectors fabricated in a single {{MOCVD}} process},\ }\href {https://doi.org/10.1088/1361-6528/ad06d1} {\bibfield  {journal} {\bibinfo  {journal} {Nanotechnology}\ }\textbf {\bibinfo {volume} {35}},\ \bibinfo {pages} {055202} (\bibinfo {year}
  {2023})}\BibitemShut {NoStop}%
\bibitem [{\citenamefont {Cheng}\ \emph {et~al.}(2018)\citenamefont {Cheng}, \citenamefont {Summerfield}, \citenamefont {Mellor}, \citenamefont {Davies}, \citenamefont {Khlobystov}, \citenamefont {Eaves}, \citenamefont {Foxon}, \citenamefont {Beton},\ and\ \citenamefont {Novikov}}]{chengHightemperatureMolecularBeam2018}%
  \BibitemOpen
  \bibfield  {author} {\bibinfo {author} {\bibfnamefont {T.~S.}\ \bibnamefont {Cheng}}, \bibinfo {author} {\bibfnamefont {A.}~\bibnamefont {Summerfield}}, \bibinfo {author} {\bibfnamefont {C.~J.}\ \bibnamefont {Mellor}}, \bibinfo {author} {\bibfnamefont {A.}~\bibnamefont {Davies}}, \bibinfo {author} {\bibfnamefont {A.~N.}\ \bibnamefont {Khlobystov}}, \bibinfo {author} {\bibfnamefont {L.}~\bibnamefont {Eaves}}, \bibinfo {author} {\bibfnamefont {C.~T.}\ \bibnamefont {Foxon}}, \bibinfo {author} {\bibfnamefont {P.~H.}\ \bibnamefont {Beton}},\ and\ \bibinfo {author} {\bibfnamefont {S.~V.}\ \bibnamefont {Novikov}},\ }\bibfield  {title} {\bibinfo {title} {High-temperature molecular beam epitaxy of hexagonal boron nitride layers},\ }\href {https://doi.org/10.1116/1.5011280} {\bibfield  {journal} {\bibinfo  {journal} {Journal of Vacuum Science \& Technology B}\ }\textbf {\bibinfo {volume} {36}},\ \bibinfo {pages} {02D103} (\bibinfo {year} {2018})}\BibitemShut {NoStop}%
\bibitem [{\citenamefont {Bonamente}(2017)}]{Bonamenta2017}%
  \BibitemOpen
  \bibfield  {author} {\bibinfo {author} {\bibfnamefont {M.}~\bibnamefont {Bonamente}},\ }\bibfield  {title} {\bibinfo {title} {Goodness of fit and parameter uncertainty},\ }in\ \href {https://doi.org/10.1007/978-1-4939-6572-4_10} {\emph {\bibinfo {booktitle} {Statistics and {{Analysis}} of {{Scientific Data}}}}},\ \bibinfo {series and number} {Graduate {{Texts}} in {{Physics}}},\ \bibinfo {editor} {edited by\ \bibinfo {editor} {\bibfnamefont {M.}~\bibnamefont {Bonamente}}}\ (\bibinfo  {publisher} {Springer},\ \bibinfo {address} {New York, NY},\ \bibinfo {year} {2017})\ pp.\ \bibinfo {pages} {177--193}\BibitemShut {NoStop}%
\bibitem [{\citenamefont {Lillie}\ \emph {et~al.}(2020)\citenamefont {Lillie}, \citenamefont {Broadway}, \citenamefont {Dontschuk}, \citenamefont {Scholten}, \citenamefont {Johnson}, \citenamefont {Wolf}, \citenamefont {Rachel}, \citenamefont {Hollenberg},\ and\ \citenamefont {Tetienne}}]{lillieLaserModulationSuperconductivity2020}%
  \BibitemOpen
  \bibfield  {author} {\bibinfo {author} {\bibfnamefont {S.~E.}\ \bibnamefont {Lillie}}, \bibinfo {author} {\bibfnamefont {D.~A.}\ \bibnamefont {Broadway}}, \bibinfo {author} {\bibfnamefont {N.}~\bibnamefont {Dontschuk}}, \bibinfo {author} {\bibfnamefont {S.~C.}\ \bibnamefont {Scholten}}, \bibinfo {author} {\bibfnamefont {B.~C.}\ \bibnamefont {Johnson}}, \bibinfo {author} {\bibfnamefont {S.}~\bibnamefont {Wolf}}, \bibinfo {author} {\bibfnamefont {S.}~\bibnamefont {Rachel}}, \bibinfo {author} {\bibfnamefont {L.~C.~L.}\ \bibnamefont {Hollenberg}},\ and\ \bibinfo {author} {\bibfnamefont {J.-P.}\ \bibnamefont {Tetienne}},\ }\bibfield  {title} {\bibinfo {title} {Laser modulation of superconductivity in a cryogenic wide-field nitrogen-vacancy microscope},\ }\href {https://doi.org/10.1021/acs.nanolett.9b05071} {\bibfield  {journal} {\bibinfo  {journal} {Nano Letters}\ }\textbf {\bibinfo {volume} {20}},\ \bibinfo {pages} {1855} (\bibinfo {year} {2020})}\BibitemShut {NoStop}%
\bibitem [{\citenamefont {Broadway}\ \emph {et~al.}(2020)\citenamefont {Broadway}, \citenamefont {Scholten}, \citenamefont {Tan}, \citenamefont {Dontschuk}, \citenamefont {Lillie}, \citenamefont {Johnson}, \citenamefont {Zheng}, \citenamefont {Wang}, \citenamefont {Oganov}, \citenamefont {Tian}, \citenamefont {Li}, \citenamefont {Lei}, \citenamefont {Wang}, \citenamefont {Hollenberg},\ and\ \citenamefont {Tetienne}}]{broadwayImagingDomainReversal2020}%
  \BibitemOpen
  \bibfield  {author} {\bibinfo {author} {\bibfnamefont {D.~A.}\ \bibnamefont {Broadway}}, \bibinfo {author} {\bibfnamefont {S.~C.}\ \bibnamefont {Scholten}}, \bibinfo {author} {\bibfnamefont {C.}~\bibnamefont {Tan}}, \bibinfo {author} {\bibfnamefont {N.}~\bibnamefont {Dontschuk}}, \bibinfo {author} {\bibfnamefont {S.~E.}\ \bibnamefont {Lillie}}, \bibinfo {author} {\bibfnamefont {B.~C.}\ \bibnamefont {Johnson}}, \bibinfo {author} {\bibfnamefont {G.}~\bibnamefont {Zheng}}, \bibinfo {author} {\bibfnamefont {Z.}~\bibnamefont {Wang}}, \bibinfo {author} {\bibfnamefont {A.~R.}\ \bibnamefont {Oganov}}, \bibinfo {author} {\bibfnamefont {S.}~\bibnamefont {Tian}}, \bibinfo {author} {\bibfnamefont {C.}~\bibnamefont {Li}}, \bibinfo {author} {\bibfnamefont {H.}~\bibnamefont {Lei}}, \bibinfo {author} {\bibfnamefont {L.}~\bibnamefont {Wang}}, \bibinfo {author} {\bibfnamefont {L.~C.~L.}\ \bibnamefont {Hollenberg}},\ and\ \bibinfo {author} {\bibfnamefont {J.-P.}\ \bibnamefont {Tetienne}},\ }\bibfield  {title} {\bibinfo
  {title} {Imaging domain reversal in an ultrathin van der waals ferromagnet},\ }\href {https://doi.org/10.1002/adma.202003314} {\bibfield  {journal} {\bibinfo  {journal} {Advanced Materials}\ }\textbf {\bibinfo {volume} {32}},\ \bibinfo {pages} {2003314} (\bibinfo {year} {2020})}\BibitemShut {NoStop}%
\bibitem [{\citenamefont {Wen}\ \emph {et~al.}(2025)\citenamefont {Wen}, \citenamefont {Pieplow}, \citenamefont {Yang}, \citenamefont {Jamshidi}, \citenamefont {Helm}, \citenamefont {Luo}, \citenamefont {Schr{\"o}der}, \citenamefont {Zhou},\ and\ \citenamefont {Berenc{\'e}n}}]{wenOpticalSpinReadout2025}%
  \BibitemOpen
  \bibfield  {author} {\bibinfo {author} {\bibfnamefont {S.}~\bibnamefont {Wen}}, \bibinfo {author} {\bibfnamefont {G.}~\bibnamefont {Pieplow}}, \bibinfo {author} {\bibfnamefont {J.}~\bibnamefont {Yang}}, \bibinfo {author} {\bibfnamefont {K.}~\bibnamefont {Jamshidi}}, \bibinfo {author} {\bibfnamefont {M.}~\bibnamefont {Helm}}, \bibinfo {author} {\bibfnamefont {J.-W.}\ \bibnamefont {Luo}}, \bibinfo {author} {\bibfnamefont {T.}~\bibnamefont {Schr{\"o}der}}, \bibinfo {author} {\bibfnamefont {S.}~\bibnamefont {Zhou}},\ and\ \bibinfo {author} {\bibfnamefont {Y.}~\bibnamefont {Berenc{\'e}n}},\ }\href {https://doi.org/10.48550/arXiv.2502.07632} {\bibinfo {title} {Optical spin readout of a silicon color center in the telecom {{L-band}}}} (\bibinfo {year} {2025}),\ \Eprint {https://arxiv.org/abs/2502.07632} {arXiv:2502.07632 [quant-ph]} \BibitemShut {NoStop}%
\bibitem [{\citenamefont {Romestain}\ and\ \citenamefont {Weisbuch}(1980)}]{romestainOpticalDetectionCyclotron1980}%
  \BibitemOpen
  \bibfield  {author} {\bibinfo {author} {\bibfnamefont {R.}~\bibnamefont {Romestain}}\ and\ \bibinfo {author} {\bibfnamefont {C.}~\bibnamefont {Weisbuch}},\ }\bibfield  {title} {\bibinfo {title} {Optical {{Detection}} of {{Cyclotron Resonance}} in {{Semiconductors}}},\ }\href {https://doi.org/10.1103/PhysRevLett.45.2067} {\bibfield  {journal} {\bibinfo  {journal} {Physical Review Letters}\ }\textbf {\bibinfo {volume} {45}},\ \bibinfo {pages} {2067} (\bibinfo {year} {1980})}\BibitemShut {NoStop}%
\bibitem [{\citenamefont {Healey}\ \emph {et~al.}(2020)\citenamefont {Healey}, \citenamefont {Stacey}, \citenamefont {Johnson}, \citenamefont {Broadway}, \citenamefont {Teraji}, \citenamefont {Simpson}, \citenamefont {Tetienne},\ and\ \citenamefont {Hollenberg}}]{healeyComparisonDifferentMethods2020}%
  \BibitemOpen
  \bibfield  {author} {\bibinfo {author} {\bibfnamefont {A.~J.}\ \bibnamefont {Healey}}, \bibinfo {author} {\bibfnamefont {A.}~\bibnamefont {Stacey}}, \bibinfo {author} {\bibfnamefont {B.~C.}\ \bibnamefont {Johnson}}, \bibinfo {author} {\bibfnamefont {D.~A.}\ \bibnamefont {Broadway}}, \bibinfo {author} {\bibfnamefont {T.}~\bibnamefont {Teraji}}, \bibinfo {author} {\bibfnamefont {D.~A.}\ \bibnamefont {Simpson}}, \bibinfo {author} {\bibfnamefont {J.-P.}\ \bibnamefont {Tetienne}},\ and\ \bibinfo {author} {\bibfnamefont {L.~C.~L.}\ \bibnamefont {Hollenberg}},\ }\bibfield  {title} {\bibinfo {title} {Comparison of different methods of nitrogen-vacancy layer formation in diamond for wide-field quantum microscopy},\ }\href {https://doi.org/10.1103/PhysRevMaterials.4.104605} {\bibfield  {journal} {\bibinfo  {journal} {Physical Review Materials}\ }\textbf {\bibinfo {volume} {4}},\ \bibinfo {pages} {104605} (\bibinfo {year} {2020})}\BibitemShut {NoStop}%
\bibitem [{\citenamefont {Nishimura}\ \emph {et~al.}(2024)\citenamefont {Nishimura}, \citenamefont {Tsukamoto}, \citenamefont {Sasaki},\ and\ \citenamefont {Kobayashi}}]{nishimuraInvestigationsOpticalAberration2024}%
  \BibitemOpen
  \bibfield  {author} {\bibinfo {author} {\bibfnamefont {S.}~\bibnamefont {Nishimura}}, \bibinfo {author} {\bibfnamefont {M.}~\bibnamefont {Tsukamoto}}, \bibinfo {author} {\bibfnamefont {K.}~\bibnamefont {Sasaki}},\ and\ \bibinfo {author} {\bibfnamefont {K.}~\bibnamefont {Kobayashi}},\ }\href {https://doi.org/10.48550/arXiv.2402.14422} {\bibinfo {title} {Investigations of optical aberration on quantum diamond microscopy toward high spatial resolution and sensitivity}} (\bibinfo {year} {2024}),\ \Eprint {https://arxiv.org/abs/2402.14422} {arXiv:2402.14422 [cond-mat, physics:physics]} \BibitemShut {NoStop}%
\bibitem [{\citenamefont {Scholten}\ \emph {et~al.}(2022)\citenamefont {Scholten}, \citenamefont {Robertson}, \citenamefont {Abrahams}, \citenamefont {Singh}, \citenamefont {Healey},\ and\ \citenamefont {Tetienne}}]{scholtenAberrationControlQuantitative2022}%
  \BibitemOpen
  \bibfield  {author} {\bibinfo {author} {\bibfnamefont {S.~C.}\ \bibnamefont {Scholten}}, \bibinfo {author} {\bibfnamefont {I.~O.}\ \bibnamefont {Robertson}}, \bibinfo {author} {\bibfnamefont {G.~J.}\ \bibnamefont {Abrahams}}, \bibinfo {author} {\bibfnamefont {P.}~\bibnamefont {Singh}}, \bibinfo {author} {\bibfnamefont {A.~J.}\ \bibnamefont {Healey}},\ and\ \bibinfo {author} {\bibfnamefont {J.-P.}\ \bibnamefont {Tetienne}},\ }\bibfield  {title} {\bibinfo {title} {Aberration control in quantitative widefield quantum microscopy},\ }\href {https://doi.org/10.1116/5.0114436} {\bibfield  {journal} {\bibinfo  {journal} {AVS Quantum Science}\ }\textbf {\bibinfo {volume} {4}},\ \bibinfo {pages} {034404} (\bibinfo {year} {2022})}\BibitemShut {NoStop}%
\bibitem [{\citenamefont {Fan}\ \emph {et~al.}(2023)\citenamefont {Fan}, \citenamefont {Myers}, \citenamefont {Sukra},\ and\ \citenamefont {Gabrielse}}]{fanMeasurementElectronMagnetic2023}%
  \BibitemOpen
  \bibfield  {author} {\bibinfo {author} {\bibfnamefont {X.}~\bibnamefont {Fan}}, \bibinfo {author} {\bibfnamefont {T.~G.}\ \bibnamefont {Myers}}, \bibinfo {author} {\bibfnamefont {B.~A.~D.}\ \bibnamefont {Sukra}},\ and\ \bibinfo {author} {\bibfnamefont {G.}~\bibnamefont {Gabrielse}},\ }\bibfield  {title} {\bibinfo {title} {Measurement of the {{Electron Magnetic Moment}}},\ }\href {https://doi.org/10.1103/PhysRevLett.130.071801} {\bibfield  {journal} {\bibinfo  {journal} {Physical Review Letters}\ }\textbf {\bibinfo {volume} {130}},\ \bibinfo {pages} {071801} (\bibinfo {year} {2023})}\BibitemShut {NoStop}%
\bibitem [{\citenamefont {Doherty}\ \emph {et~al.}(2012)\citenamefont {Doherty}, \citenamefont {Dolde}, \citenamefont {Fedder}, \citenamefont {Jelezko}, \citenamefont {Wrachtrup}, \citenamefont {Manson},\ and\ \citenamefont {Hollenberg}}]{dohertyTheoryGroundstateSpin2012}%
  \BibitemOpen
  \bibfield  {author} {\bibinfo {author} {\bibfnamefont {M.~W.}\ \bibnamefont {Doherty}}, \bibinfo {author} {\bibfnamefont {F.}~\bibnamefont {Dolde}}, \bibinfo {author} {\bibfnamefont {H.}~\bibnamefont {Fedder}}, \bibinfo {author} {\bibfnamefont {F.}~\bibnamefont {Jelezko}}, \bibinfo {author} {\bibfnamefont {J.}~\bibnamefont {Wrachtrup}}, \bibinfo {author} {\bibfnamefont {N.~B.}\ \bibnamefont {Manson}},\ and\ \bibinfo {author} {\bibfnamefont {L.~C.~L.}\ \bibnamefont {Hollenberg}},\ }\bibfield  {title} {\bibinfo {title} {Theory of the ground-state spin of the {{NV}} - center in diamond},\ }\href {https://doi.org/10.1103/physrevb.85.205203} {\bibfield  {journal} {\bibinfo  {journal} {Physical Review B}\ }\textbf {\bibinfo {volume} {85}},\ \bibinfo {pages} {205203} (\bibinfo {year} {2012})}\BibitemShut {NoStop}%
\bibitem [{\citenamefont {Fairley}\ \emph {et~al.}(2021)\citenamefont {Fairley}, \citenamefont {Fernandez}, \citenamefont {Richard-Plouet}, \citenamefont {{Guillot-Deudon}}, \citenamefont {Walton}, \citenamefont {Smith}, \citenamefont {Flahaut}, \citenamefont {Greiner}, \citenamefont {Biesinger}, \citenamefont {Tougaard}, \citenamefont {Morgan},\ and\ \citenamefont {Baltrusaitis}}]{fairleySystematicCollaborativeApproach2021}%
  \BibitemOpen
  \bibfield  {author} {\bibinfo {author} {\bibfnamefont {N.}~\bibnamefont {Fairley}}, \bibinfo {author} {\bibfnamefont {V.}~\bibnamefont {Fernandez}}, \bibinfo {author} {\bibfnamefont {M.}~\bibnamefont {Richard-Plouet}}, \bibinfo {author} {\bibfnamefont {C.}~\bibnamefont {{Guillot-Deudon}}}, \bibinfo {author} {\bibfnamefont {J.}~\bibnamefont {Walton}}, \bibinfo {author} {\bibfnamefont {E.}~\bibnamefont {Smith}}, \bibinfo {author} {\bibfnamefont {D.}~\bibnamefont {Flahaut}}, \bibinfo {author} {\bibfnamefont {M.}~\bibnamefont {Greiner}}, \bibinfo {author} {\bibfnamefont {M.}~\bibnamefont {Biesinger}}, \bibinfo {author} {\bibfnamefont {S.}~\bibnamefont {Tougaard}}, \bibinfo {author} {\bibfnamefont {D.}~\bibnamefont {Morgan}},\ and\ \bibinfo {author} {\bibfnamefont {J.}~\bibnamefont {Baltrusaitis}},\ }\bibfield  {title} {\bibinfo {title} {Systematic and collaborative approach to problem solving using {{X-ray}} photoelectron spectroscopy},\ }\href {https://doi.org/10.1016/j.apsadv.2021.100112} {\bibfield
  {journal} {\bibinfo  {journal} {Applied Surface Science Advances}\ }\textbf {\bibinfo {volume} {5}},\ \bibinfo {pages} {100112} (\bibinfo {year} {2021})}\BibitemShut {NoStop}%
\bibitem [{\citenamefont {Rollings}\ \emph {et~al.}(2006)\citenamefont {Rollings}, \citenamefont {Gweon}, \citenamefont {Zhou}, \citenamefont {Mun}, \citenamefont {McChesney}, \citenamefont {Hussain}, \citenamefont {Fedorov}, \citenamefont {First}, \citenamefont {{de Heer}},\ and\ \citenamefont {Lanzara}}]{rollingsSynthesisCharacterizationAtomically2006}%
  \BibitemOpen
  \bibfield  {author} {\bibinfo {author} {\bibfnamefont {E.}~\bibnamefont {Rollings}}, \bibinfo {author} {\bibfnamefont {G.~H.}\ \bibnamefont {Gweon}}, \bibinfo {author} {\bibfnamefont {S.~Y.}\ \bibnamefont {Zhou}}, \bibinfo {author} {\bibfnamefont {B.~S.}\ \bibnamefont {Mun}}, \bibinfo {author} {\bibfnamefont {J.~L.}\ \bibnamefont {McChesney}}, \bibinfo {author} {\bibfnamefont {B.~S.}\ \bibnamefont {Hussain}}, \bibinfo {author} {\bibfnamefont {A.~V.}\ \bibnamefont {Fedorov}}, \bibinfo {author} {\bibfnamefont {P.~N.}\ \bibnamefont {First}}, \bibinfo {author} {\bibfnamefont {W.~A.}\ \bibnamefont {{de Heer}}},\ and\ \bibinfo {author} {\bibfnamefont {A.}~\bibnamefont {Lanzara}},\ }\bibfield  {title} {\bibinfo {title} {Synthesis and characterization of atomically thin graphite films on a silicon carbide substrate},\ }\href {https://doi.org/10.1016/j.jpcs.2006.05.010} {\bibfield  {journal} {\bibinfo  {journal} {Journal of Physics and Chemistry of Solids}\ }\bibinfo {series} {{{SMEC}} 2005},\ \textbf {\bibinfo
  {volume} {67}},\ \bibinfo {pages} {2172} (\bibinfo {year} {2006})}\BibitemShut {NoStop}%
\bibitem [{\citenamefont {Fujiwara}(2007)}]{fujiwaraSpectroscopicEllipsometryPrinciples2007}%
  \BibitemOpen
  \bibfield  {author} {\bibinfo {author} {\bibfnamefont {H.}~\bibnamefont {Fujiwara}},\ }\href@noop {} {\emph {\bibinfo {title} {Spectroscopic {{Ellipsometry}}: {{Principles}} and {{Applications}}}}}\ (\bibinfo  {publisher} {John Wiley \& Sons},\ \bibinfo {year} {2007})\BibitemShut {NoStop}%
\end{thebibliography}%

\clearpage
\onecolumngrid

\begin{center}
\textbf{\large Supplementary Information for the manuscript ``Optically detected magnetic resonance of wafer-scale hexagonal boron nitride thin films''}
\end{center}

\setcounter{equation}{0}
\setcounter{section}{0}
\setcounter{figure}{0}
\setcounter{table}{0}
\setcounter{page}{1}
\makeatletter
\renewcommand{\theequation}{S\arabic{equation}}
\renewcommand{\thefigure}{S\arabic{figure}}

\section{Sample thickness quantification}

\subsection{MBE sample thickness}

\begin{figure*}[ht!]
\centering
\includegraphics[width=0.99\textwidth]{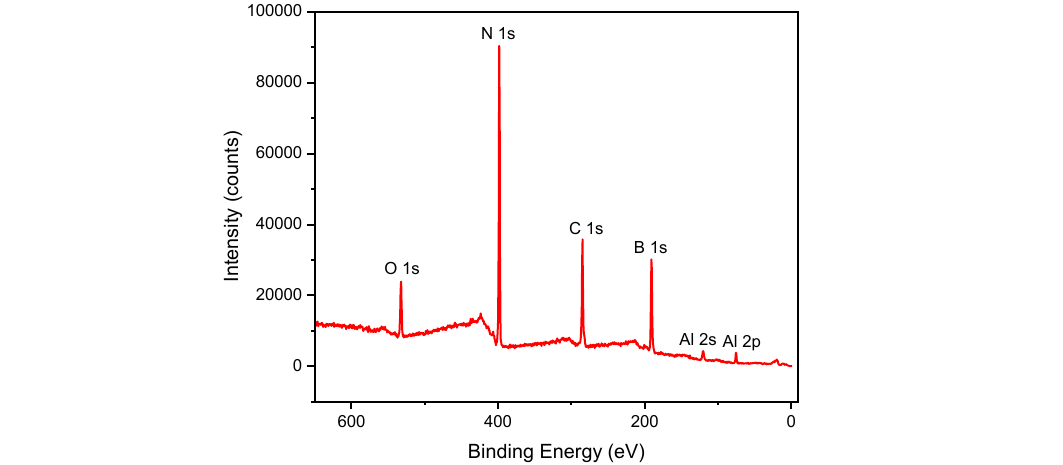}
\caption{\textbf{XPS of the MBE-grown hBN epilayer on a sapphire substrate.}
The relative intensities of the nitrogen (N\,1s) and boron (B\,1s) core level peaks to the aluminium peaks (Al 2s and Al 2p) can be analysed to determine the layer thickness. }
\label{sifig:0I}
\end{figure*}

For the sample grown by MBE the hBN epilayer thickness was determined by x-ray photoelectron spectroscopy (XPS) and confirmed by variable angle spectroscopic ellipsometry (VASE). 
The XPS spectra (Fig.~\ref{sifig:0I}) were recorded on a Kratos Axis Ultra DLD instrument with a monochromated Al K$\alpha$ X-ray source ($h\nu = 1486.7$\,eV). 
Survey spectra were acquired at a pass energy of 160\,eV, with charge compensation provide by flooding the surface with 3.5\,eV electrons. 
Data were measured at three points on the sample and analysed using the CasaXPS software package~\cite{fairleySystematicCollaborativeApproach2021}. 
The binding energy scale was calibrated by a rigid of the spectra to align the C 1s core level of adventitious carbon to 285\,eV.

Photoelectrons emitted from the sapphire substrate are attenuated as they scatter inelastically in the hBN overlayer. 
This results in variations of the Al 2p peak intensity relative to N 1s and B 1s peaks depending on the hBN thickness. The average layer thickness ($t$) can be calculated from the intensity ratio of B 1s or N 1s ($I_\mathrm{hBN}$) to the Al 2p intensity ($I_\mathrm{sub}$) from Eq.~\ref{eq:s1}~\cite{rollingsSynthesisCharacterizationAtomically2006},
\begin{equation}\label{eq:s1}
    \frac{I_\mathrm{hBN}}{I_\mathrm{sub}} = \frac{T(E_\mathrm{hBN}) \rho' C_\mathrm{hBN} \lambda'(E_\mathrm{hBN}) \big [ 1 - \mathrm{e}^{\frac{-t}{\lambda '(E_\mathrm{hBN})}} \big ]}{T(E_\mathrm{sub}) \rho C_\mathrm{sub} \lambda(E_\mathrm{sub}) \mathrm{e}^{\frac{-t}{\lambda (E_\mathrm{sub})}}}
\end{equation}
where $T$ is the transmission function of the electron energy analyser at the specified photoelectron energy, $C$ is the photoionisation cross-section of the relevant core level electrons, $\lambda$ is the inelastic mean free mpth (IMFP) of electrons at the specified photoelectron kinetic energy for the relevant material and $\rho$ is the atomic density.
Primed quantities (e.g.\ $\lambda '$) refer to the hBN overlayer.

The exponential factors account for attenuation of photoelectrons as they pass through the hBN overlayer.
The coefficients account for the different `efficiencies' at which photoelectrons are generated, scattered, and detected in an XPS measurement.
Solving Eq.~\ref{eq:s1}, accounting for both $I_\mathrm{N1s} / I_\mathrm{Al2p}$ and $I_\mathrm{B 1s} / I_\mathrm{Al 2p}$, gives an hBN layer thickness of $9.2\pm0.8$\,nm.

Variable-angle spectroscopic ellipsometry (VASE) measures the change in the polarisation state of light reflected from the sample. 
The ratio of the complex reflection coefficient for p-polarized light to the complex reflection coefficient for s-polarized light, denoted $\rho$, can be expressed as  
\begin{equation}\label{eq:s2}
    \rho = \tan(\Psi \cdot \mathrm{e}^{\mathrm{i} \Delta} )
\end{equation}
where $\Psi$ and $\Delta$ are the parameters (expressed as angles) returned by the ellipsometry measurement at each wavelength~\cite{fujiwaraSpectroscopicEllipsometryPrinciples2007}. 
An optical model, that includes the thickness of the epilayer and the optical properties of the materials is then constructed and fitted to the experimental data using a nonlinear fitting algorithm.

\begin{figure*}[ht!]
\centering
\includegraphics[width=0.99\textwidth]{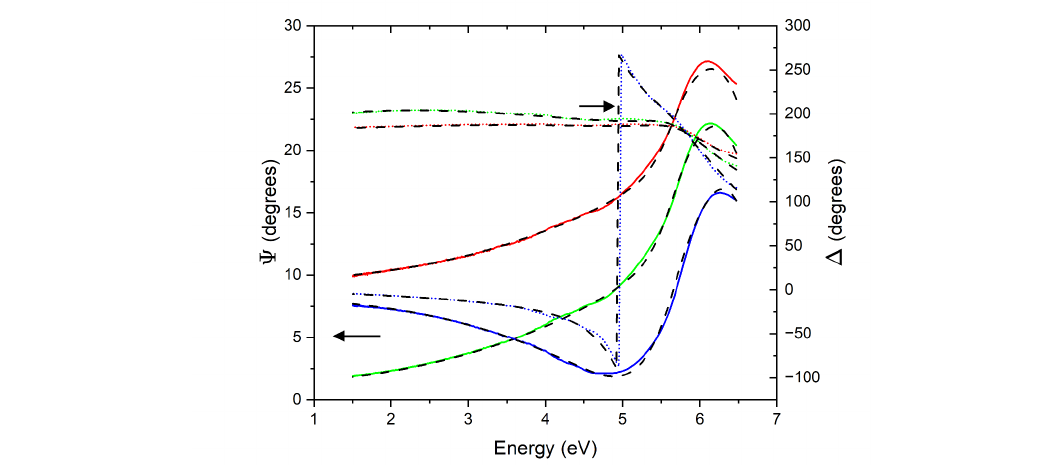}
\caption{\textbf{VASE spectroscopy of MBE-grown hBN sample.}
Typical spectroscopic ellipsometry parameters $\Psi$ (solid line) and $\Delta$ (dashed line) shown from a photon energy of 1.5 eV to 6.5 eV for three angles of incidence (55$\degree$: red; 60$\degree$: green; and 65$\degree$: blue) along with the model fitted to the data (black dashed line). }
\label{sifig:0II}
\end{figure*}

\begin{figure*}[ht!]
\centering
\includegraphics[width=0.99\textwidth]{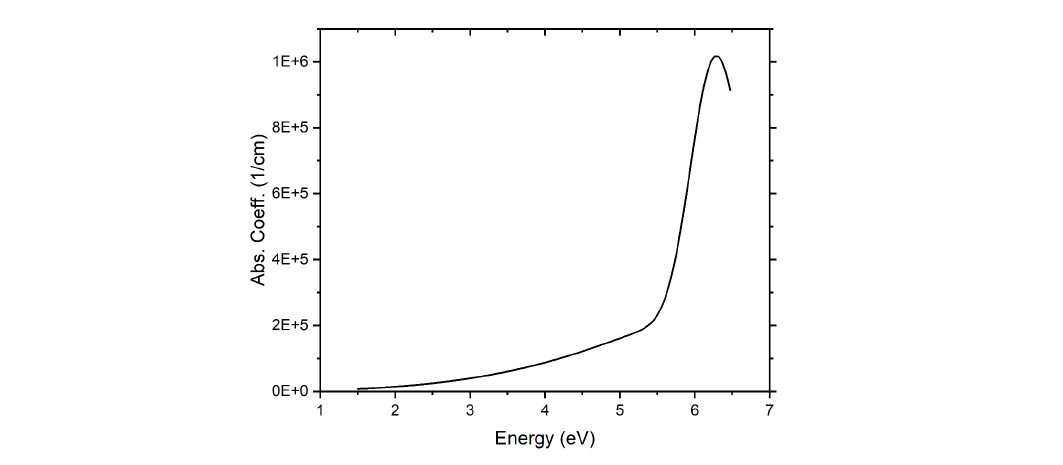}
\caption{\textbf{Absorption coefficient for the hBN epilayer predicted by the optical model.}}
\label{sifig:0III}
\end{figure*}

When the hBN films are thinner than 10nm, the simultaneous determination of the refractive index and the film thickness from a single spectroscopic ellipsometry measurement becomes unreliable due to cross-correlations between fitting parameters.
For the sample reported here we made nine VASE measurements on a $3\times3$ grid (2\,mm grid spacing) using focus probes (with a 0.2\,mm minor axis of the elliptical spot) on a J.A.\ Woollam Inc.\ M2000-DI spectroscopic ellipsometer at incident angles of 55$\degree$, 60$\degree$ and 65$\degree$. 
The data were modelled assuming that the same optical model holds for all nine points and the thickness of the epilayer is allowed to vary between points i.e.\ the fitting variance is mainly reflected in the variance of the epilayer thickness (Fig.~\ref{sifig:0II}). 
This type of analysis is termed multi-sample analysis (MSA) in the CompleteEase v6.73 analysis software.  
This approach reduces the cross-correlations however these are still significant leading to an epilayer thickness of $8\pm3$\,nm, a value that is consistent with the XPS value but with a higher degree of uncertainty. 
The high level of intentional carbon doping in this sample results in a measurable optical absorption at photon energies below the 6\,eV band gap of hBN (Fig.~\ref{sifig:0III}).

\subsection{AFM sample thickness verification} 

All but the ANU MOCVD sample had thicknesses characterised via optical means - X-ray reflectivity for the CVD samples, XPS and VASE for the MBE sample as noted above, and optical reflectivity for the UWarsaw MOCVD samples.
To confirm and verify these thicknesses, we etch each sample and measure their thickness via AFM.
The same method was used to measure the thickness of the ANU MOCVD sample.
The process is as follows.

\begin{figure*}[ht!]
\centering
\includegraphics[width=0.99\textwidth]{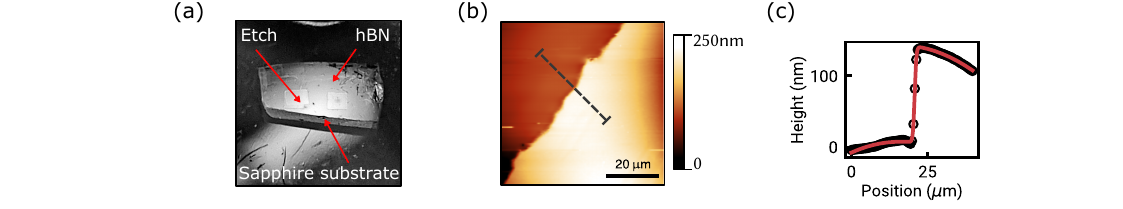}
\caption{\textbf{AFM thickness verification of hBN samples.}
(a)~Photograph of MOCVD hBN sample grown on sapphire substrate, with etched regions. 
(b, c)~AFM image (b) and profile across the edge (c), fit with a Boltzmann bent step to extract the sample thickness.}
\label{sifig:1}
\end{figure*}

Each sample is reactive ion etched (RIE; Oxford Instrument PLASMALAB100 ICP380) through a shadow mask in SF$_6$ plasma (IF-100, ICP-1000) for two minutes to completely remove the undesired hBN regions [Fig.~\ref{sifig:1}(a)].
An atomic force microscope (AFM) image is then acquired over the etch edge [Fig.~\ref{sifig:1}(b)] and a line profile used to extract the edge height [Fig.~\ref{sifig:1}(c)].
AFM images were acquired using an Oxford Instruments Asylum Research MFP-3D Infinity AFM instrument in AC mode with a Budget Sensors Tap300AL-G probe. 
The images were collected at a scan rate of 0.5\,Hz.
We characterised the etch process on a bare sapphire substrate and found a 3\,nm step, suggesting a very slow sapphire etch rate, although we note that hBN overlayer may affect the sapphire etch rate (due to potential effects on the sapphire surface during growth).
Additionally, the step height varied slightly across the sample, so we cautiously estimate the thickness relative uncertainty at 5\%.
The measured thicknesses are shown in Table~\ref{sitab:1}, showing general agreement aside from a couple of outliers.
We note that MOCVD-B was too thick to fully etch in our two minute RIE, though the measurement confirms the large thickness.
The MBE sample is the greatest outlier, showing an order of magnitude greater AFM thickness than the optical measurement.
Given that XPS and VASE measurements of the MBE thickness agree, we take the optical thickness as more correct for our analysis.
Finally, we note that the AFM thickness does not account for the non-uniform density of the films (which may have some porosity), and so is a less accurate measure of sensor performance per crystal volume compared with the optical measurements.

\begin{table*}[ht!]
\begin{tabularx}{0.6\textwidth}{ c c Y Y } \toprule
 \multirow{2}{*}{Sample Name} & \multirow{2}{*}{Source} & AFM step & Optical step \\ 
                              &                         & (nm) & (nm) \\ \midrule
 MOCVD-A1 & UWarsaw  & 44   & 39   \\ 
 MOCVD-A4 & UWarsaw  & 30    & 29   \\
 MOCVD-B  & UWarsaw  & 1450 & 2000 \\ 
 MOCVD-D  &     ANU  & 40   & -   \\ 
 MBE      &     UoN  & 86   & 9    \\
 CVD-1    & AIXTRON  & 14   & 9    \\ \bottomrule
\end{tabularx}
\caption{\textbf{Comparison of AFM and optically measured hBN film thicknesses.}
} 
\label{sitab:1}
\end{table*}

\section{PL spectra for all samples}

\begin{figure*}[ht!]
\centering
\includegraphics[width=0.99\textwidth]{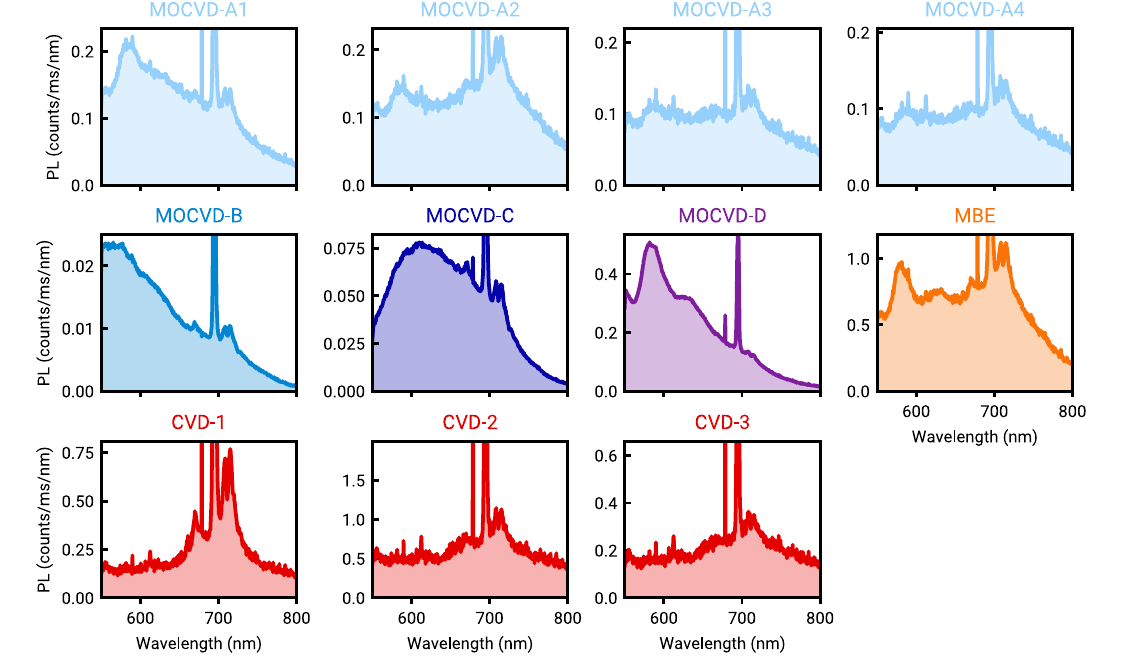}
\caption{\textbf{PL spectra of survey series.}
Acquired under 532\,nm excitation.}
\label{sifig:2}
\end{figure*}

\begin{figure*}[ht!]
\centering
\includegraphics[width=0.99\textwidth]{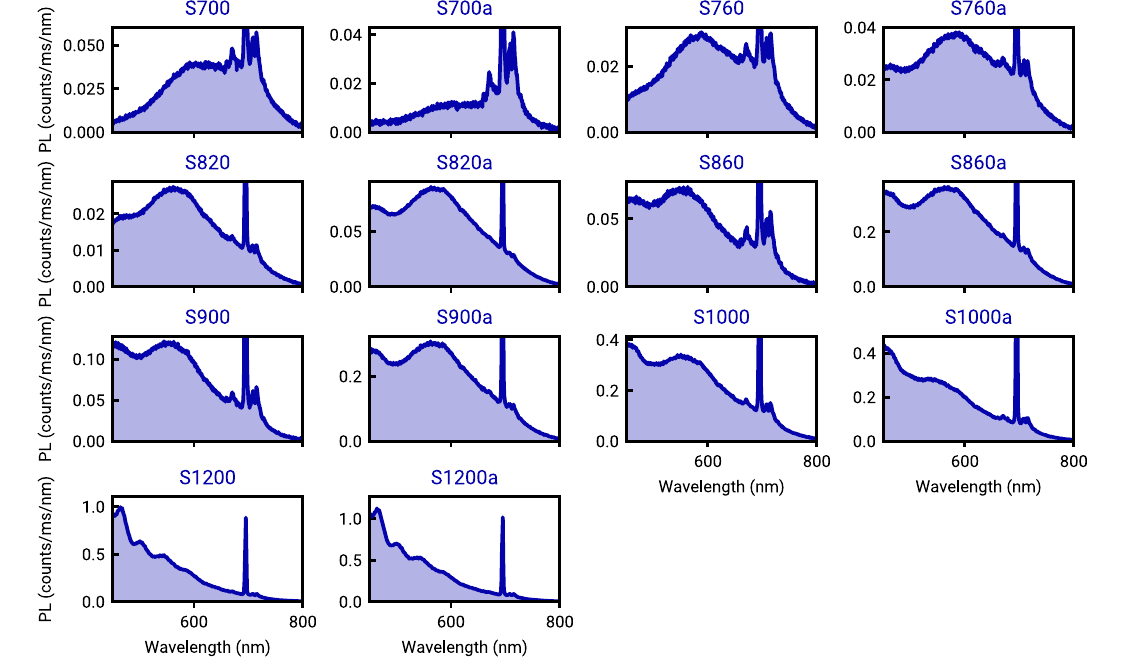}
\caption{\textbf{PL spectra of MOCVD growth tem perature series.} Acquired under 405\,nm excitation.}
\label{sifig:3}
\end{figure*}

Figures \ref{sifig:2},\ref{sifig:3} display measured PL spectra for all of the samples in this paper.
Here the spectra are shown without masking the Cr$^{3+}$ optical transition in the substrate sapphire (sharp features near 690\,nm).

\end{document}